\documentclass[a4paper,11pt]{article}

\usepackage[free-standing-units]{siunitx}
\usepackage{hepunits}
\usepackage{hepnicenames}
\usepackage{fullpage}
\usepackage{amsmath}
\usepackage{amssymb}
\usepackage{graphicx}
\usepackage{hyperref}
\usepackage{xspace}
\DeclareGraphicsExtensions{.pdf}

\newcommand{\Wo}{\ensuremath{\mathrm{W}}\xspace}%
\newcommand{\Zo}{\ensuremath{\mathrm{Z}}\xspace}%
\newcommand{\Vo}{\ensuremath{\mathrm{V}}\xspace}%
\newcommand{\MW}{\ensuremath{m_{\Wo}}\xspace}%
\newcommand{\MZ}{\ensuremath{m_{\Zo}}\xspace}%
\newcommand{\Zprime}{\Zo^\prime\xspace}%
\newcommand{\llbar}{\ensuremath{\ell^+\ell^-}\xspace}%
\renewcommand{\Pq}{\mathrm{q}}%
\renewcommand{\Paq}{\mathrm{\overline{q}}}%
\renewcommand{\Pg}{\mathrm{g}}%
\newcommand{\qqbar}{\ensuremath{\Pq\Paq}}%
\newcommand{\pp}{\ensuremath{\mathrm{pp}}\xspace}%
\newcommand{\ppVll}{\ensuremath{\pp\to\Vo^{(*)}\to\llbar}\xspace}%
\newcommand{\ppVln}{\ensuremath{\pp\to\Wo^{(*)}\to\ell\nu}\xspace}%
\newcommand{\qqVgvirt}{\qqbar\to\Vo(\Pg)\xspace}%
\newcommand{\qgVq}{\Pq\Pg\to\Vo\Pq\xspace}%
\newcommand{\qbargVqbar}{\Paq g\to \Vo\Paq\xspace}%
\newcommand{\chiq}{\chi_\mathrm{q}\xspace}%
\newcommand{\chil}{\chi_\ell\xspace}%
\newcommand{\rhon}{\rho_{\mathrm{n}}\xspace}%
\newcommand{\tq}{T^3(\mathrm{q}_{\chi_\mathrm{q}})\xspace}%
\newcommand{\tl}{T^3(\ell_{\chi_\ell})\xspace}%
\newcommand{\tf}{T^3(\mathrm{f}_{\chi_\mathrm{f}})\xspace}%
\newcommand{\yq}{Y(\mathrm{q}_{\chi_\mathrm{q}})\xspace}%
\newcommand{\yl}{Y(\ell_{\chi_\ell})\xspace}%
\newcommand{\yf}{Y(\mathrm{f}_{\chi_\mathrm{f}})\xspace}%
\newcommand{\Qq}{Q(q)\xspace}%
\newcommand{\Ql}{Q(\ell)\xspace}%
\newcommand{\AFB}{A_{FB}\xspace}%
\newcommand{\abinv}{\,\text{ab}^{-1}\xspace}
\newcommand{\Wpar}{{\ensuremath{\cal W}}\xspace}%
\newcommand{\Ypar}{{\ensuremath{\cal Y}}\xspace}%

\begin{document}

\title{An optimal strategy for experimental searches in high-mass dilepton production at the LHC}
\author{J. Alcaraz Maestre, O. M. Carretero \\ 
	\small Centro de Investigaciones Energéticas, Medioambientales y Tecnológicas (CIEMAT), Madrid
	\small \footnote{The authors acknowledge support from the research project PID2023-147706NB-I00 funded by the Spanish MICIU/AEI /10.13039/501100011033 and FEDER/EU.}
}
\date{}

\maketitle

\begin{abstract}
We present an optimal method for studying new physics effects in high-mass dilepton production at the LHC. This method relies on a reweighting procedure that is precise to the next-to-leading order in quantum chromodynamics. The only computationally expensive input required is a standard model $\ppVll$ sample, with $\Vo=\Zo,\gamma^*$, generated at the same level of precision and with sufficient statistical coverage of the phase space at high invariant masses. The proposed method is suitable for any $\qqbar\llbar$ contact-interaction scenario, regardless of the assumed underlying structure and couplings. It is also ideal for searching for medium-to-large-width new neutral resonances produced in the s-channel, as changes to the input values for mass, coupling, or width can be easily implemented using the same initial generated samples.
\end{abstract}

\section{Introduction} 

Current and future experimental searches for new physics in high-mass dilepton final states
at the LHC require a precise understanding of the underlying process, $\ppVll$, to within
one percent or better, where $\Vo\equiv \Zo,\gamma^*$. At high luminosities, this requires
the use of very large simulated event samples, as well as quantum chromodynamics (QCD) and
electroweak (EWK) corrections that go beyond leading-order perturbation theory.
Eventually, a substantial number of beyond the standard model (BSM) processes will need to
be explored with this level of precision. However, producing so many large, precise
simulated event samples is unrealistic. An interesting step in this direction was reported
in Ref.~\cite{Andrea-main}, in the context of compositeness scenarios characterized by
non-zero values of the \Wpar and \Ypar oblique parameters~\cite{WY_definition}. Their
strategy involves minor ad hoc modifications to the SM POWHEG-BOX~\cite{POWHEG-Z}. Here,
we extend that strategy and propose a reweighting procedure at the generator level that is
much more convenient for an experimental analysis. This new approach is well-suited for any
BSM model with $\qqbar\llbar$ contact interactions and can be easily extended to any BSM
scenario with $\Zprime$ resonances of medium/large width. 

\section{First steps in the strategy \label{sec:Andrea-method}}

Ref.~\cite{Andrea-main} shows that next-to-leading (NLO) QCD corrections and, to a large
extent, NLO EWK corrections can be easily taken into account in SM+BSM scenarios involving
$\llbar$ and $\ell\nu$ final states by treating separately the generation of events with
different fermion chiralities. One can assume that chirality is a good quantum number
because the massless limit for all fermions is an accurate approximation for the Drell-Yan
process at high energies. These higher-order corrections can be easily calculated in the
case of final states mediated by the $\Wo$ boson ($\ppVln$), since only left-handed
fermions are involved in the process. These charged-current corrections have already been
used in LHC experiments to set limits on the $\Wpar$ parameter~\cite{CMS-Run2-ln}. However,
the situation is more complicated in the $\ppVll$ case. 

Let us denote by $\chiq$ the chirality of the initial state quark and by $\chil$ the
chirality of the final state lepton. At Born level, the (SM+BSM)/SM event weight for a
given combination of $\chiq$ and $\chil$ is simply a global factor that depends only on the
dilepton invariant mass. The key point is that no additional modifications to the weight
are needed at NLO. In this scheme, the generation sequence would be as follows: a) generate
events independently for each $\chiq\chil$ case; b) reweight each event according to a
weight that depends only on the dilepton invariant mass, $\chiq$ and $\chil$; c) add all
reweighted $(\chiq,\chil)$ contributions according to their respective SM cross sections. 

We will demonstrate how this procedure can be significantly simplified by avoiding the
generation of separate samples for each chirality combination. Using ordinary NLO SM
samples, which are averaged over helicities, we will be able to reweight events in-situ. 

Let us first introduce the BSM model explored in Ref.~\cite{Andrea-main}. It 
considers the following $\qqbar\llbar$ contact interaction terms:

\begin{eqnarray}
	{\cal O'}_{2W} & = & {\displaystyle -\frac{g^2~\Wpar~\tq~\tl}{\MW^2}~[\Paq_{\chiq} \gamma_\mu \Pq_{\chiq}]~[\overline{\ell}_{\chil} \gamma^\mu \ell_{\chil}] + \ldots} \\
	{\cal O'}_{2Y} & = & {\displaystyle -\frac{g^{\prime 2}~\Ypar~\yq~\yl}{\MW^2}~[\Paq_{\chiq} \gamma_\mu \Pq_{\chiq}]~[\overline{\ell}_{\chil} \gamma^\mu \ell_{\chil}] + \ldots}
\end{eqnarray}

In the previous expressions $\Pq_{\chiq}$ and $\ell_{\chil}$ denote quark and lepton fields
of either left or right-handed chirality. The coupling constants $g$ and $g'$ are those
associated with the $SU(2)_L$ and $U(1)_Y$ groups of the SM, respectively. The third
component of isospin of the fermion $\mathrm{f}_{\chi_\mathrm{f}}$, which is non-zero only
for left-handed fields, is denoted by $\tf$, and $\yf$ is its hypercharge. The oblique
parameters \Wpar and \Ypar~\cite{WY_definition}, which are zero at tree level in the SM,
are used to quantify BSM effects. The mass of the W boson, $\MW$, is arbitrarily introduced
to provide the right six-dimensionality to the contact term. From a more physical point of
view the natural scale of new physics should be $\Lambda$, with $\Lambda\gg \sqrt{s}$, i.e.
${\cal O}(\Wpar/\MW^2) = {\cal O}(\Ypar/\MW^2) = {\cal O}(c_{\chiq\chil}/\Lambda^2)$ with
an associated coupling $c_{\chiq\chil}$ of magnitude ${\cal O}(1)$:

\begin{eqnarray}
	-\frac{g^2~\Wpar~\tq~\tl}{\MW^2}, -\frac{g^{\prime 2}~\Ypar~\yq~\yl}{\MW^2} & ~~\longrightarrow~~ & \frac{c_{\chiq\chil}}{\Lambda^2}
	\label{eq:CI}
\end{eqnarray}

For a given chirality combination, the event weight factor $\rhon$ reads~\cite{Andrea-main}:

\begin{eqnarray}
	\rhon(s,\Wpar,\Ypar;\chiq,\chil) & = & \left| \frac{C^0_\mathrm{SM}(s;\chiq,\chil)+C^0_\mathrm{BSM}(s,\Wpar,\Ypar;\chiq,\chil)}{C^0_\mathrm{SM}(s;\chiq,\chil)} \right|^2
	\label{eq:old-reweight}
\end{eqnarray}

\begin{eqnarray}
	C^0_\mathrm{SM}(s;\chiq,\chil) & = & {\displaystyle \frac{g^2\tq\tl + g'^2\yq\yl - e^2\Qq\Ql~\frac{\MZ^2}{s}}{s-\MZ^2}}\;, \\
	C^0_\mathrm{BSM}(s,\Wpar,\Ypar;\chiq,\chil) & = & {\displaystyle \frac{- g^2\tq\tl~\Wpar - g'^2\yq\yl~\Ypar}{\MW^2}}\;,
\end{eqnarray}

\noindent where $\Qq$ and $\Ql$ are the electric charges of quark and lepton in positron 
units, respectively.

\section{Angular distributions at NLO accuracy}

\begin{figure}[htbp]
\centering
\includegraphics[width=0.80\linewidth]{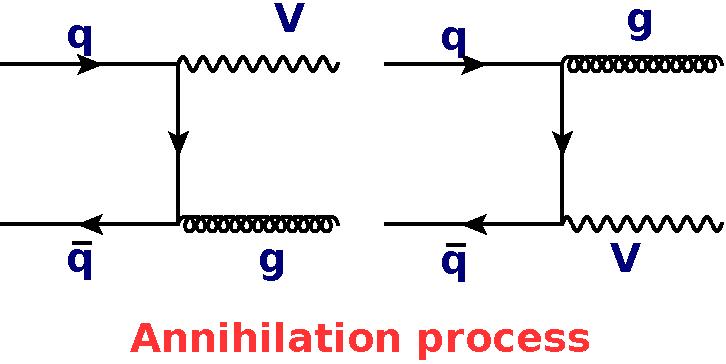}\\
\includegraphics[width=0.80\linewidth]{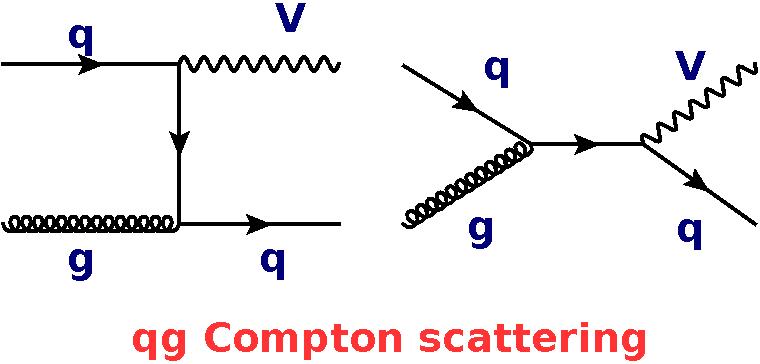}\\
\includegraphics[width=0.80\linewidth]{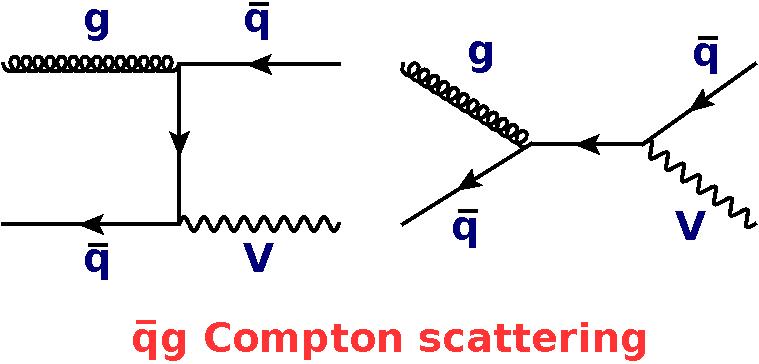}

\caption{Diagrams for the V + 1 jet process. \label{fig:diagrams}}

\end{figure}

At NLO, processes involving the emission of one additional quark, antiquark, or gluon are present. In the $\Vo$ rest frame, the normalized lepton angular distribution up to NLO precision is given by~\cite{Chaichian}: 

\begin{eqnarray}
\frac{d\rho}{d\Omega_{\ell}} & = & \frac{3}{16\pi}~\left[1+\frac{|\vec{p_1}|^2~\cos^2\theta_{1l}+|\vec{p_2}|^2\cos^2\theta_{2l}}{|\vec{p_1}|^2+|\vec{p_2}|^2} + 2\xi~\frac{|\vec{p_1}|^2~\cos\theta_{1l}-|\vec{p_2}|^2~\cos\theta_{2l}}{|\vec{p_1}|^2+|\vec{p_2}|^2} \right]\;, \label{eq:angular-distribution}
\end{eqnarray}

\noindent 
where $d\Omega_{\ell} = d\phi_\ell d(\cos\theta_\ell)$ is the differential solid angle of
the lepton. The $\vec{p_1}$ vector denotes either the three-momentum of the colliding quark
in $\qqVgvirt$ and $\qgVq$ subprocesses, or the three-momentum of the final-state antiquark
in the $\qbargVqbar$ subprocess. Correspondingly, $\vec{p_2}$ is the three-momentum of
either the colliding antiquark or the final-state quark, depending on the subprocess under
consideration. Figure~\ref{fig:diagrams} shows this convention more graphically:
$\vec{p_1}$ always refers to the incoming fermion arrow and $\vec{p_2}$ to the outgoing
fermion arrow. Similarly, $\theta_{1l}$ ($\theta_{2l}$) is the angle between the
$\vec{p_1}$ ($\vec{p_2}$) direction and the lepton direction. Due to rotational invariance,
Eq.~\ref{eq:angular-distribution} is valid for any center-of-mass frame choice of the V
system. 

In the above expression the physical parameter $\xi$ is non-zero due to the fact that
left-handed and right-handed couplings differ in weak interactions. It satisfies the
condition $|\xi|\le 1$ and is proportional to the forward-backward asymmetry $\AFB$ of the
associated hard-scattering process. For neutral gauge bosons
$\xi=\frac{4}{3}~\AFB(M_{\Vo^{(*)}})$.

\section{Optimal reweighting procedure}

Besides the $\rhon$ reweighting factor discussed in Section~\ref{sec:Andrea-method}, the differential cross section at NLO contains additional factors that still depend on chirality. The interesting feature is that this remaining dependence is fully absorbed into the lepton differential angular distribution introduced in the previous section. For left-left ($LL$) and right-right ($RR$) combinations of chirality the distributions follow Equation~\ref{eq:angular-distribution} with $\xi=+1$: 

\begin{eqnarray}
\frac{d\rho^{LL}}{d\Omega_{\ell}} = \frac{d\rho^{RR}}{d\Omega_{\ell}} & = & \frac{3}{16\pi}~\left[\frac{|\vec{p_1}|^2~(1+\cos\theta_{1l})^2+|\vec{p_2}|^2(1-\cos\theta_{2l})^2}{|\vec{p_1}|^2+|\vec{p_2}|^2} \right]\;, \label{eq:angular-distribution-LL}
\end{eqnarray}

\noindent whereas for left-right ($LR$) and right-left ($RL$) combinations the 
distributions follow Equation~\ref{eq:angular-distribution} with $\xi=-1$:

\begin{eqnarray}
\frac{d\rho^{LR}}{d\Omega_{\ell}} = \frac{d\rho^{RL}}{d\Omega_{\ell}} & = & \frac{3}{16\pi}~\left[\frac{|\vec{p_1}|^2~(1-\cos\theta_{1l})^2+|\vec{p_2}|^2(1+\cos\theta_{2l})^2}{|\vec{p_1}|^2+|\vec{p_2}|^2} \right]\;. \label{eq:angular-distribution-LR}
\end{eqnarray}

Note that these distributions simplify significantly at Born level because $\cos\theta_{2l}=-\cos\theta_{1l}$. These simplified forms are also good approximations for low transverse momentum values of the dilepton system when the Collins-Soper reference frame~\cite{CollinsSoper} is used (corrections appear at order ${\cal O}(p^2_T(\Vo)/M^2_{\Vo})$):

\begin{eqnarray}
\frac{d\rho^{LL}}{d\Omega_{\ell}} = \frac{d\rho^{RR}}{d\Omega_{\ell}} & \approx & \frac{3}{16\pi}~\left(1+\cos\theta_{CS}\right)^2\;, \label{eq:angular-distribution-LL-simple} \\
\frac{d\rho^{LR}}{d\Omega_{\ell}} = \frac{d\rho^{RL}}{d\Omega_{\ell}} & \approx & \frac{3}{16\pi}~\left(1-\cos\theta_{CS}\right)^2\;, \label{eq:angular-distribution-LR-simple}
\end{eqnarray}

\noindent where $\theta_{CS}$ denotes the Collins-Soper polar angle~\footnote{The
Collins-Soper polar angle is given by: ${\displaystyle \cos\theta_{CS} =
\frac{2(p_z^{\ell^-}E^{\ell^+} - p_z^{\ell^+}E^{\ell-})}{M\sqrt{M^2+p_T^2}}}$, where
$E,p_z$ denote energy and longitudinal momentum, and $M,p_T$ are the mass and transverse
momentum of the $\llbar$ system. The positive Z axis is given by the quark particle in the
beam.}. Nevertheless, we will use the exact expressions from
Eqs.~\ref{eq:angular-distribution-LL}-\ref{eq:angular-distribution-LR} in the reweighting
at generator level, to ensure NLO precision in the full phase space. 

The expression for the event reweighting factor can now be easily deduced. This factor is
used to transform a set of ordinary SM events generated at NLO QCD into a set of SM+BSM
events at the same order of precision:

\begin{eqnarray}
	\rhon(s,\Wpar,\Ypar) & = & \frac
	{\sum\limits_{\chiq,\chil} {\displaystyle \frac{d\rho^{\chiq\chil}}{d\Omega_{\ell}} \left|(C^0_\mathrm{SM}(s;\chiq,\chil)+C^0_\mathrm{BSM}(s,\Wpar,\Ypar;\chiq,\chil)\right|^2}}
	{\sum\limits_{\chiq,\chil} {\displaystyle \frac{d\rho^{\chiq\chil}}{d\Omega_{\ell}} \left|C^0_\mathrm{SM}(s;\chiq,\chil)\right|^2}}\;,
	\label{eq:new-reweight}
\end{eqnarray}

\noindent with $\chiq,\chil \in \{L,R\}$.

\section{Validation and quantification of BSM effects via oblique parameters.}

We adopted the procedure outlined in Ref.~\cite{Andrea-main}, and modified a
POWHEG-Z~\cite{POWHEG-Z} version accordingly. This modified version provides an easy way to
generate BSM samples with different values of the oblique parameters \Wpar and \Ypar. To
validate and explore the new implementation, these BSM reference samples are then compared
with the corresponding reweighted samples obtained from an ordinary SM POWHEG generation
using the weights from Eq.~\ref{eq:new-reweight}. 

All POWHEG samples, consisting of one million events each, were generated at a
center-of-mass energy of $\sqrt{s}=13~\TeV$ with dilepton invariant masses in two different
intervals: $1.4~\TeV < M_{\llbar} > 2.3~\TeV$ and $2.3~\TeV < M_{\llbar} > 6.0~\TeV$. This
was done to ensure sufficient statistics in the explored mass range, given the exponential
decrease of the differential cross section with mass. The NNPDF4.0 NNLO PDF set was used
and an integrated luminosity of ${\cal L}=1~\abinv$ was chosen for the distributions in the
comparisons. Even if a change of PDF set may slightly change the normalization and the
shapes of the distributions, it does not modify neither the size of the differences nor the
conclusions reached in the comparison. 

We examine the two key observables that are most affected by any potentially wrong
calculation of the weights. They are taken from the hard scattering level information, and
are: 1) the invariant mass of the dilepton system and 2) a realistic ``proxy'' of the 
Collins-Soper polar angle of the lepton, $\cos\theta^*_{CS}$, which assumes a reference 
system where the direction of the initial quark is assigned to the flight direction of the 
initial parton with the largest momentum.

\begin{figure}[htbp]
\centering

\includegraphics[width=0.45\linewidth]{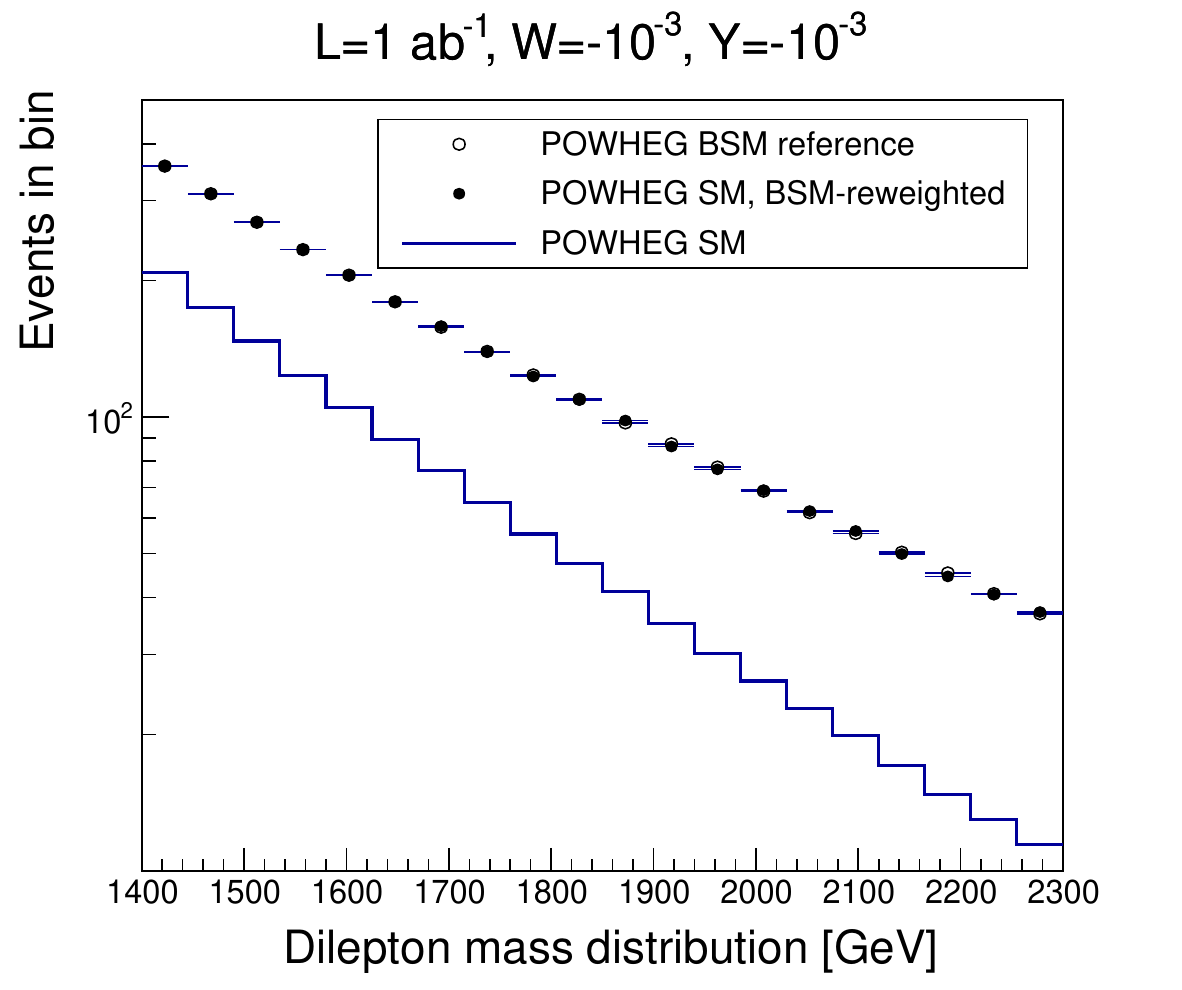}
\includegraphics[width=0.45\linewidth]{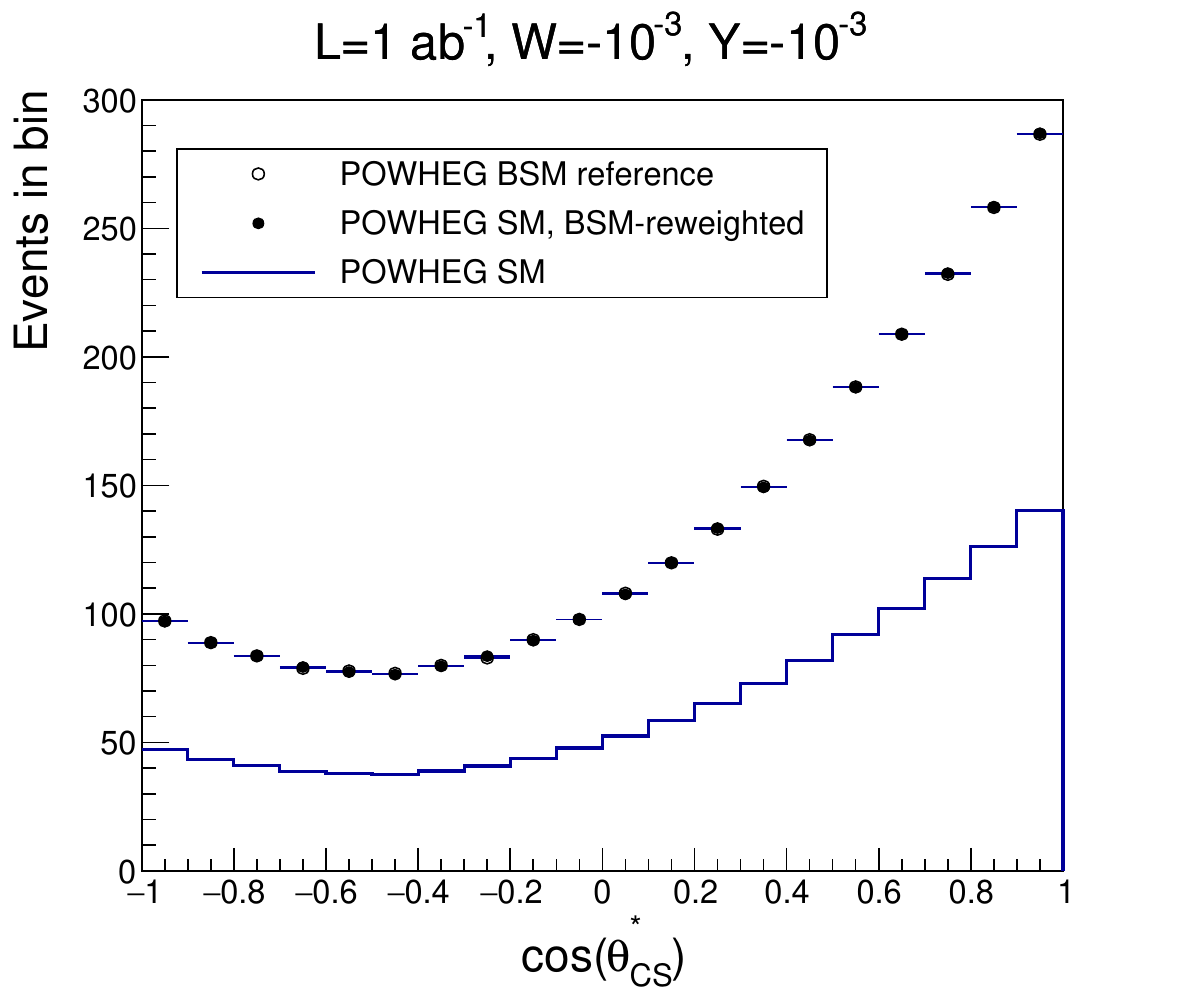}\\
\includegraphics[width=0.45\linewidth]{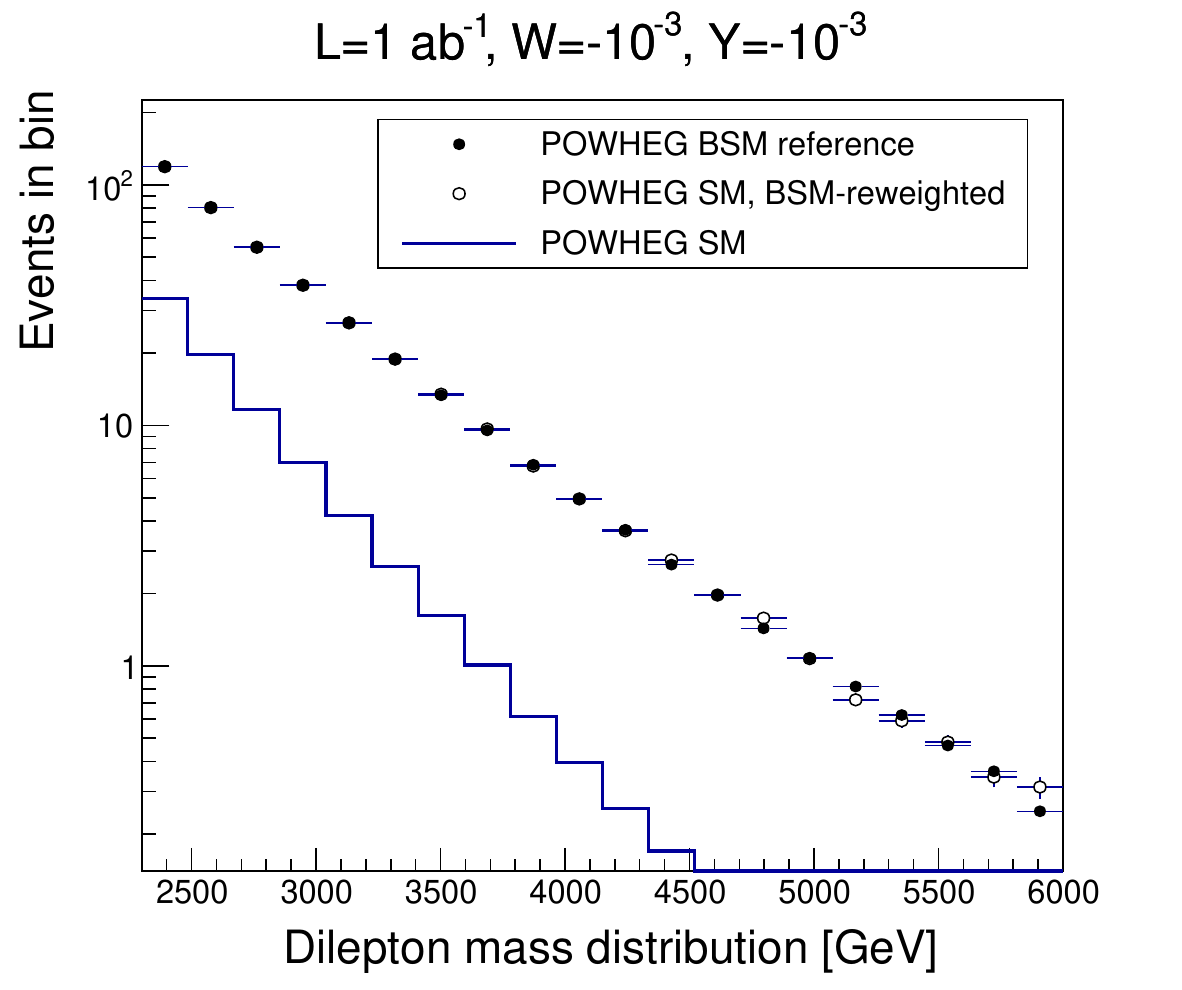}
\includegraphics[width=0.45\linewidth]{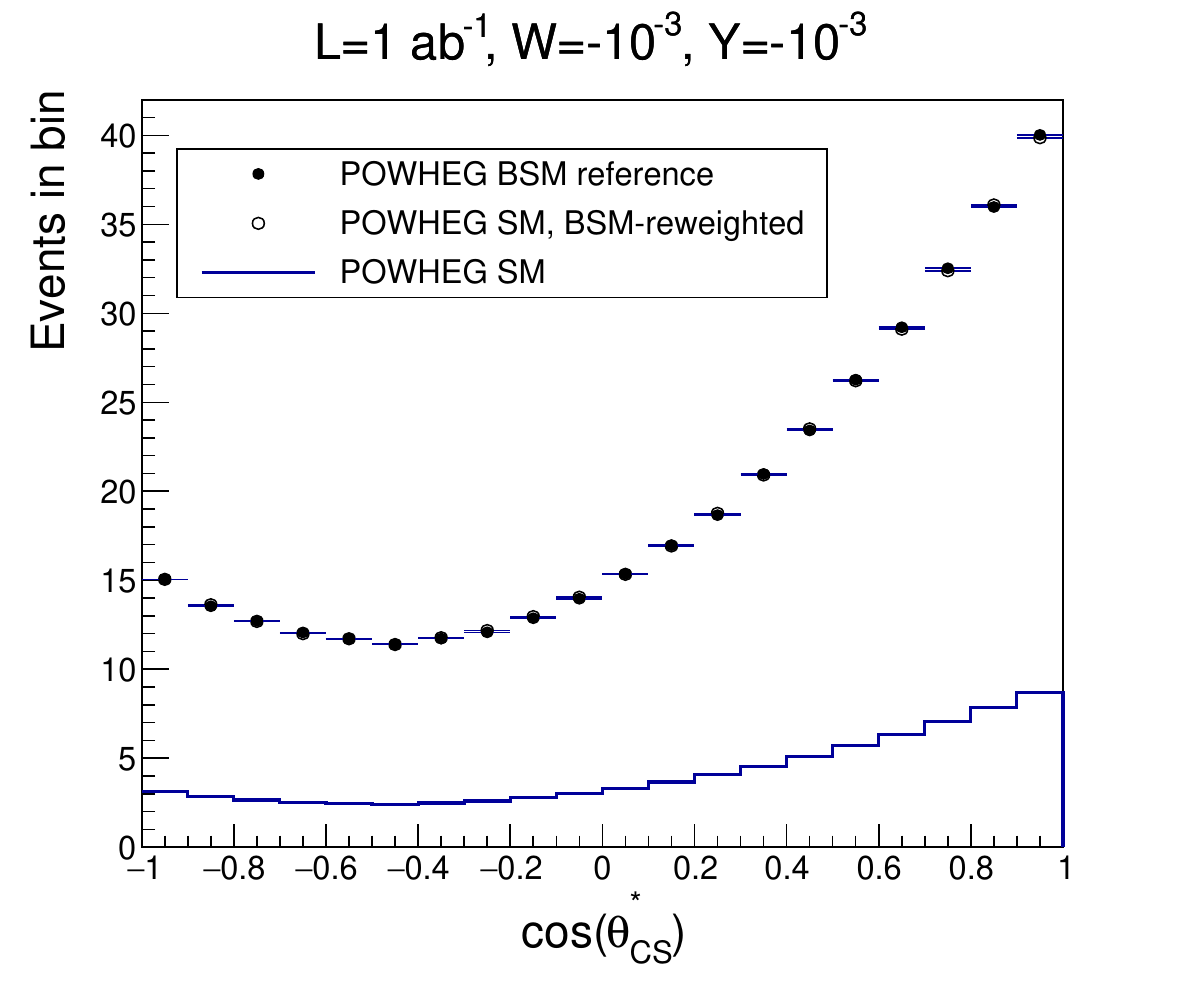}

\caption{Mass and angular comparisons between a BSM POWHEG sample generated with
$\Wpar=\Ypar=-10^{-3}$ and the corresponding result from the new proposed reweighting on a
pure SM sample inclusive in helicities. The samples have been generated with invariant
masses of the dilepton system in the intervals $1.4 < M_{\llbar} < 2.3~\TeV$ and $2.3 <
M_{\llbar} < 6.0~\TeV$ at $\sqrt{s}=13~\TeV$, for an integrated luminosity of 
${\cal L}=1~\abinv$. The SM expectations are also shown to illustrate the sensitivity to 
BSM effects.}

\label{fig:BSM-mm}

\end{figure}

Figure~\ref{fig:BSM-mm} shows the consistency checks for a BSM scenario with
$\Wpar=\Ypar=-10^{-3}$. This scenario provides a significant cross section increase at high
dilepton invariant masses - it is therefore implicitly excluded by current LHC
observations. The white-dotted distributions are extracted from the reference POWHEG
sample. The black-dotted distribution corresponds to our reweighted sample, using
Equation~\ref{eq:new-reweight} on an event by event basis. There is agreement between
reference and reweighted distributions within statistical uncertainties, both in terms of
normalization and shape. The SM expectations are also shown to illustrate the sensitivity
to BSM effects.

\begin{figure}[htbp]
\centering

\includegraphics[width=0.45\linewidth]{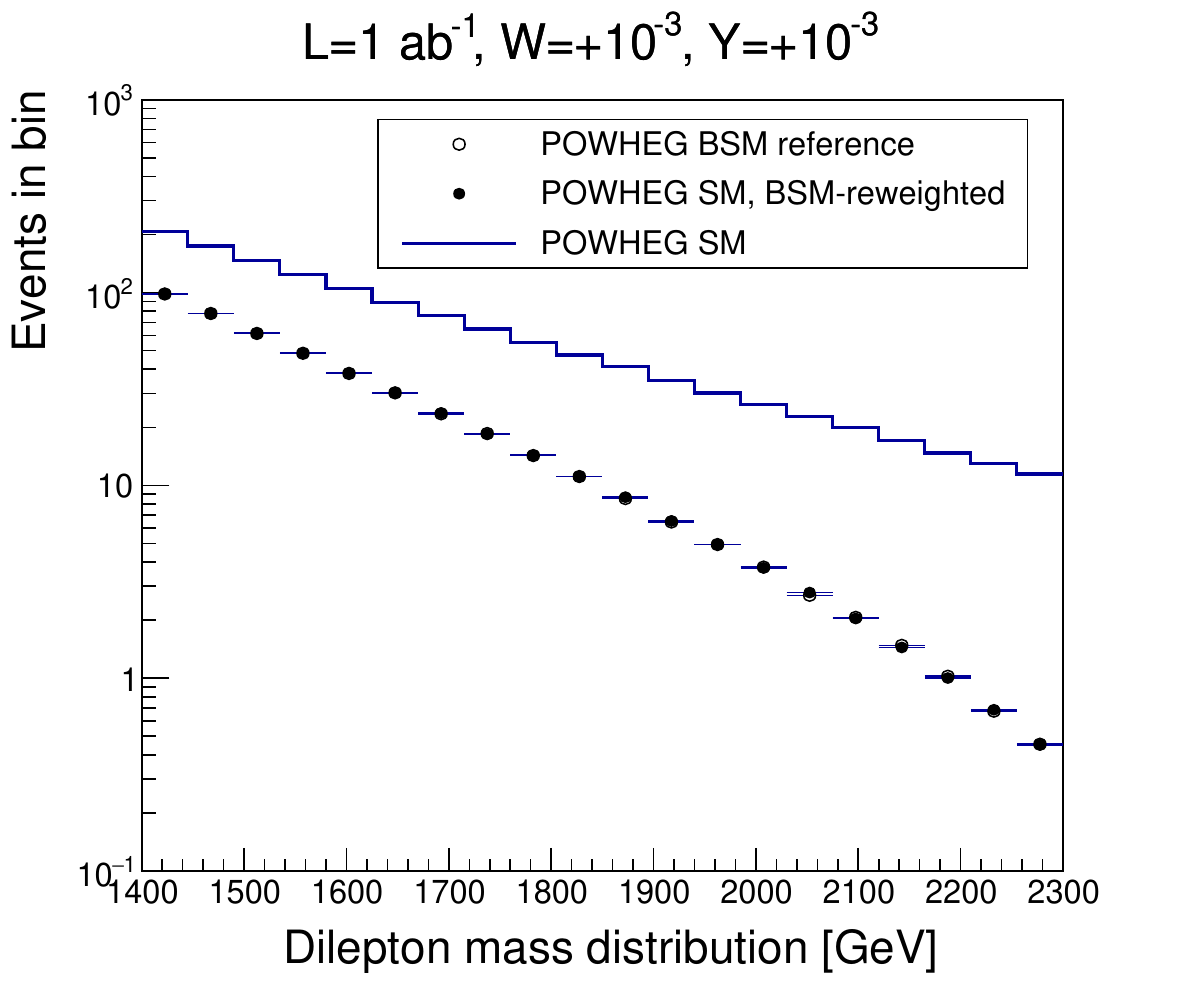}
\includegraphics[width=0.45\linewidth]{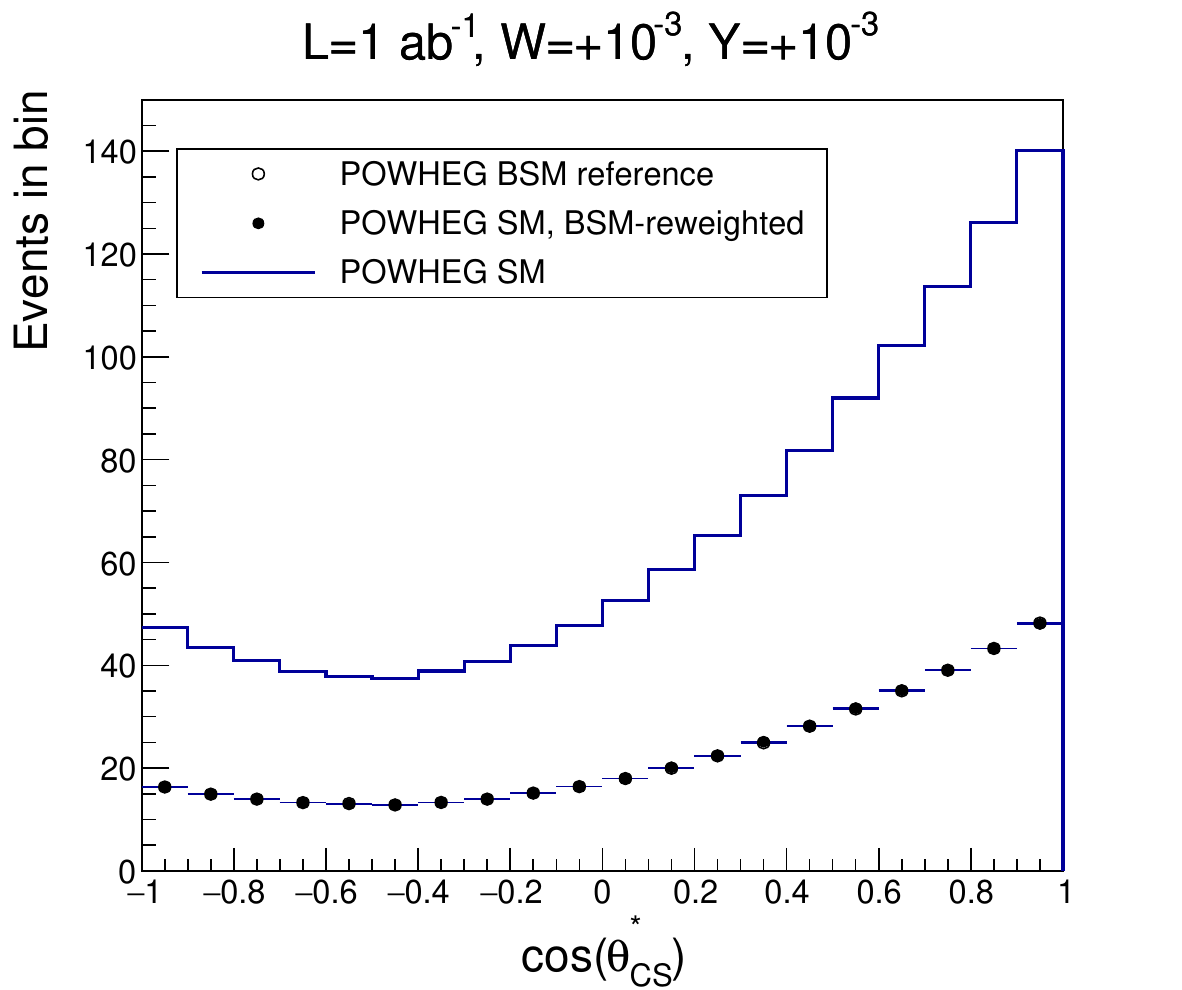}\\
\includegraphics[width=0.45\linewidth]{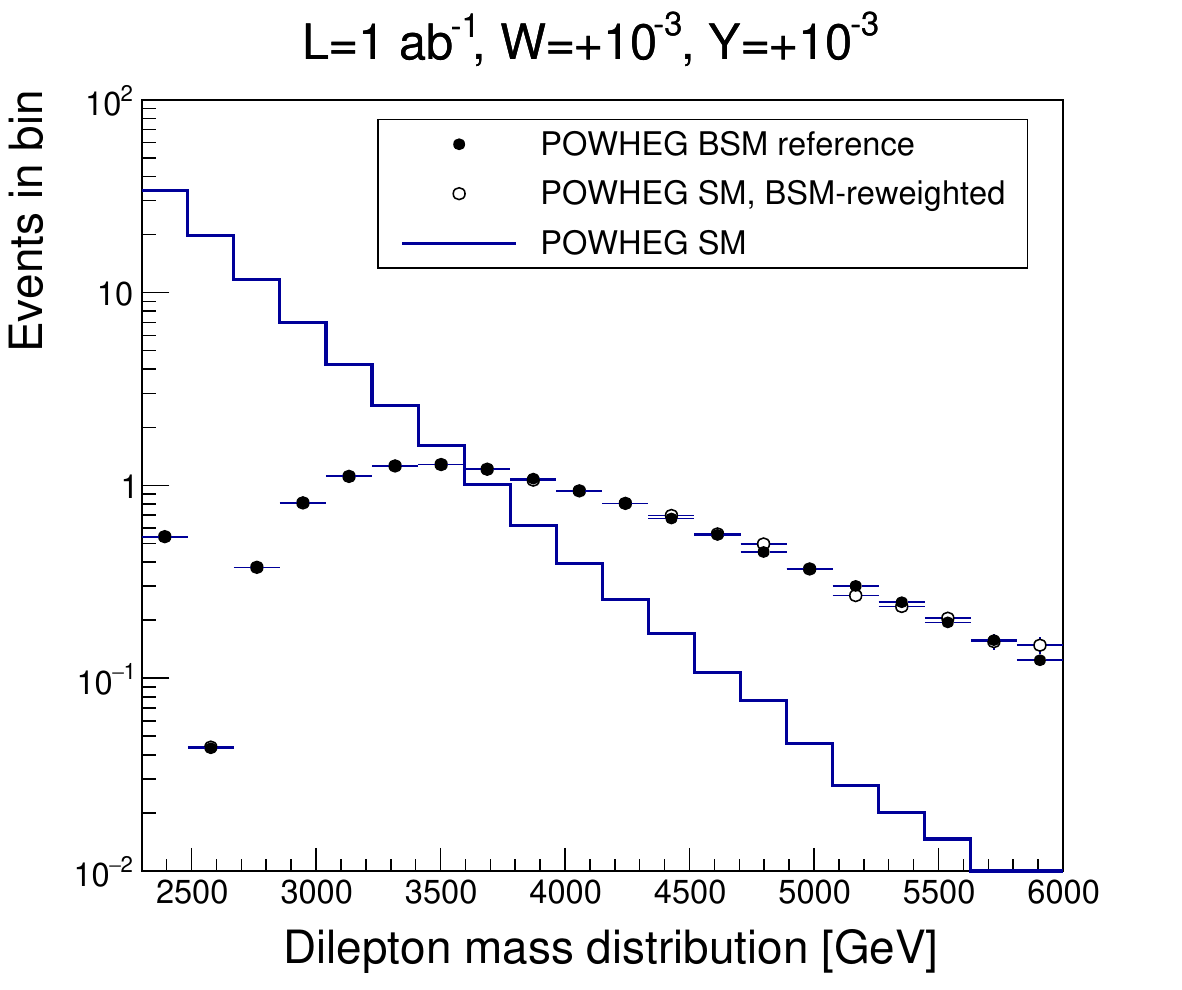}
\includegraphics[width=0.45\linewidth]{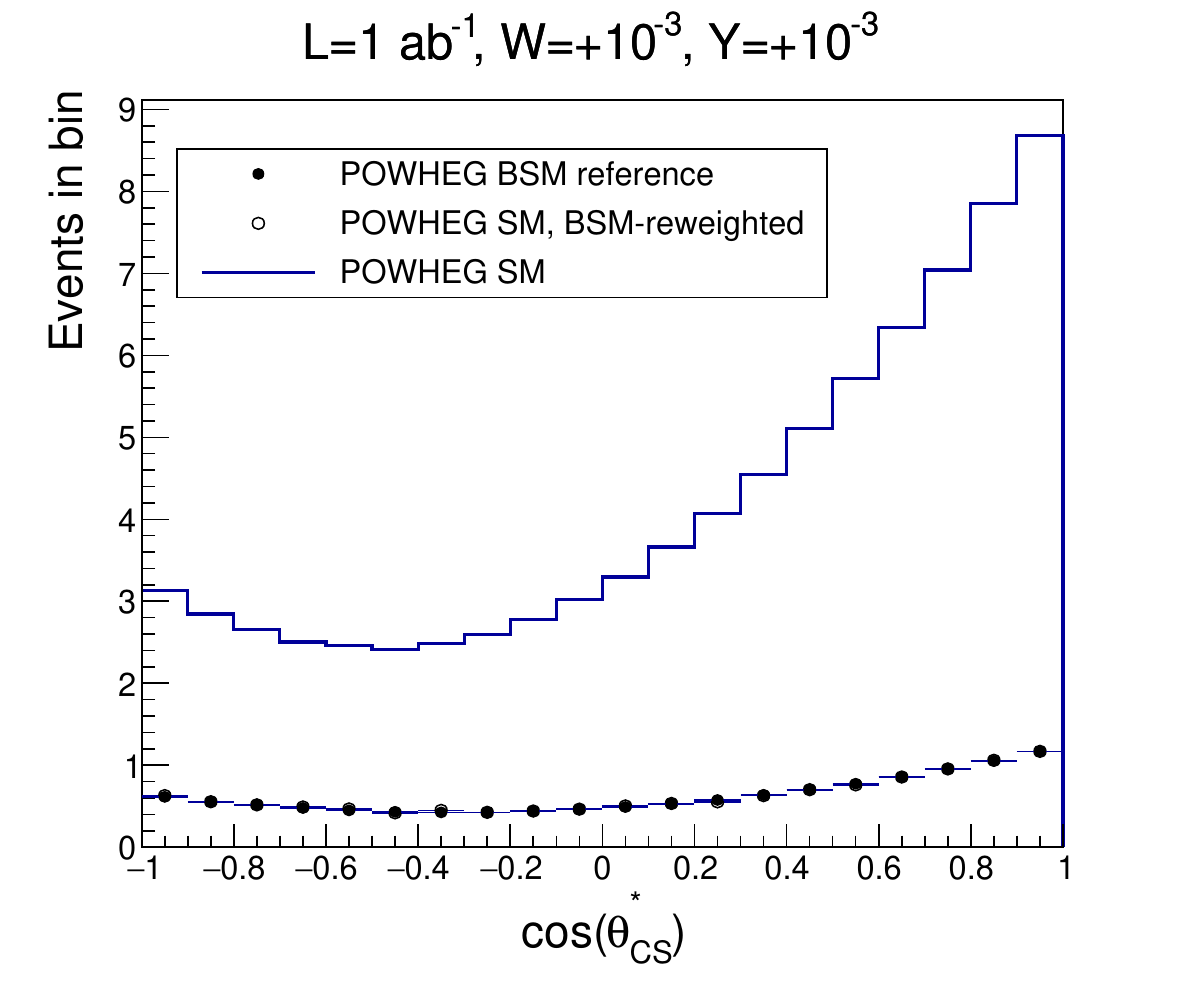}

\caption{Mass and angular comparisons between a BSM POWHEG sample generated with
$\Wpar=\Ypar=+10^{-3}$ and the corresponding result from the new proposed reweighting on a
pure SM sample inclusive in helicities. The samples have been generated with invariant
masses of the dilepton system in the intervals $1.4 < M_{\llbar} < 2.3~\TeV$ and $2.3 <
M_{\llbar} < 6.0~\TeV$ at $\sqrt{s}=13~\TeV$, for an integrated luminosity of 
${\cal L}=1~\abinv$. The SM expectations are also shown to illustrate the sensitivity to 
BSM effects. Note the significant decrease in cross section around $M_{\llbar}\approx
\MW/\sqrt{\Wpar}\equiv\MW/\sqrt{\Ypar}$, which is expected due to the similarities between
BSM and SM structures at high mass.} 

\label{fig:BSM-pp}

\end{figure}

\begin{figure}[htbp]
\centering

\includegraphics[width=0.45\linewidth]{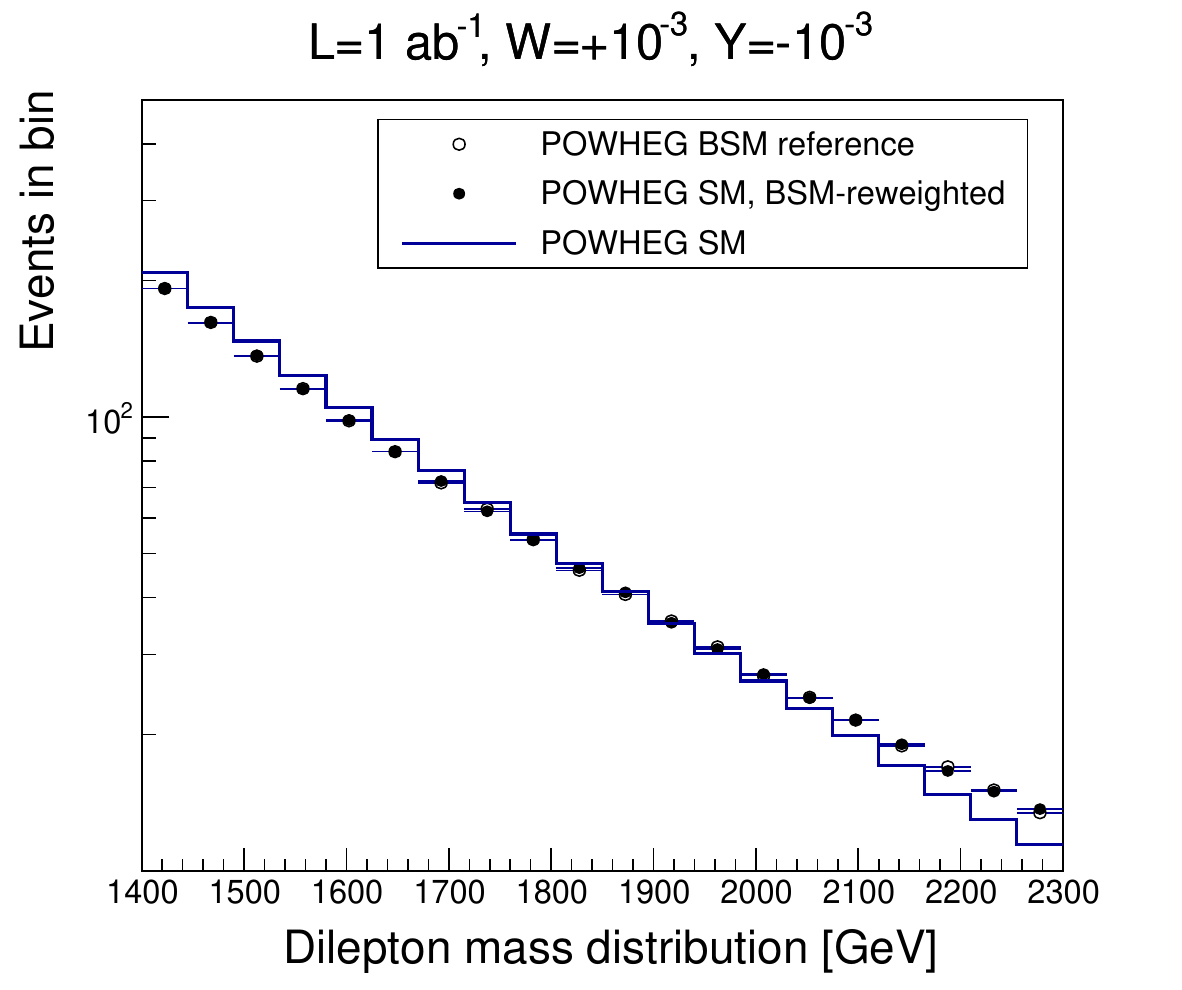}
\includegraphics[width=0.45\linewidth]{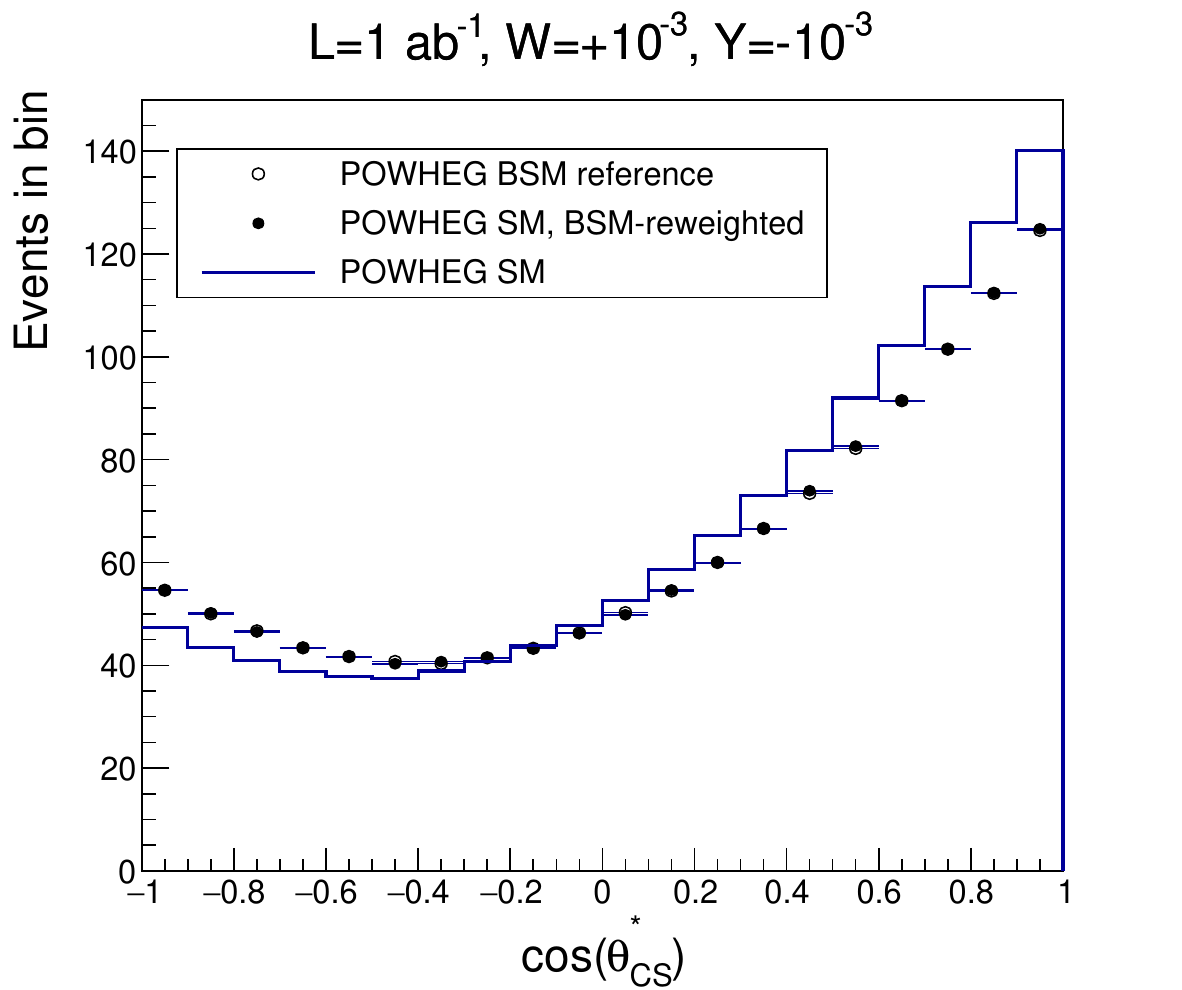}\\
\includegraphics[width=0.45\linewidth]{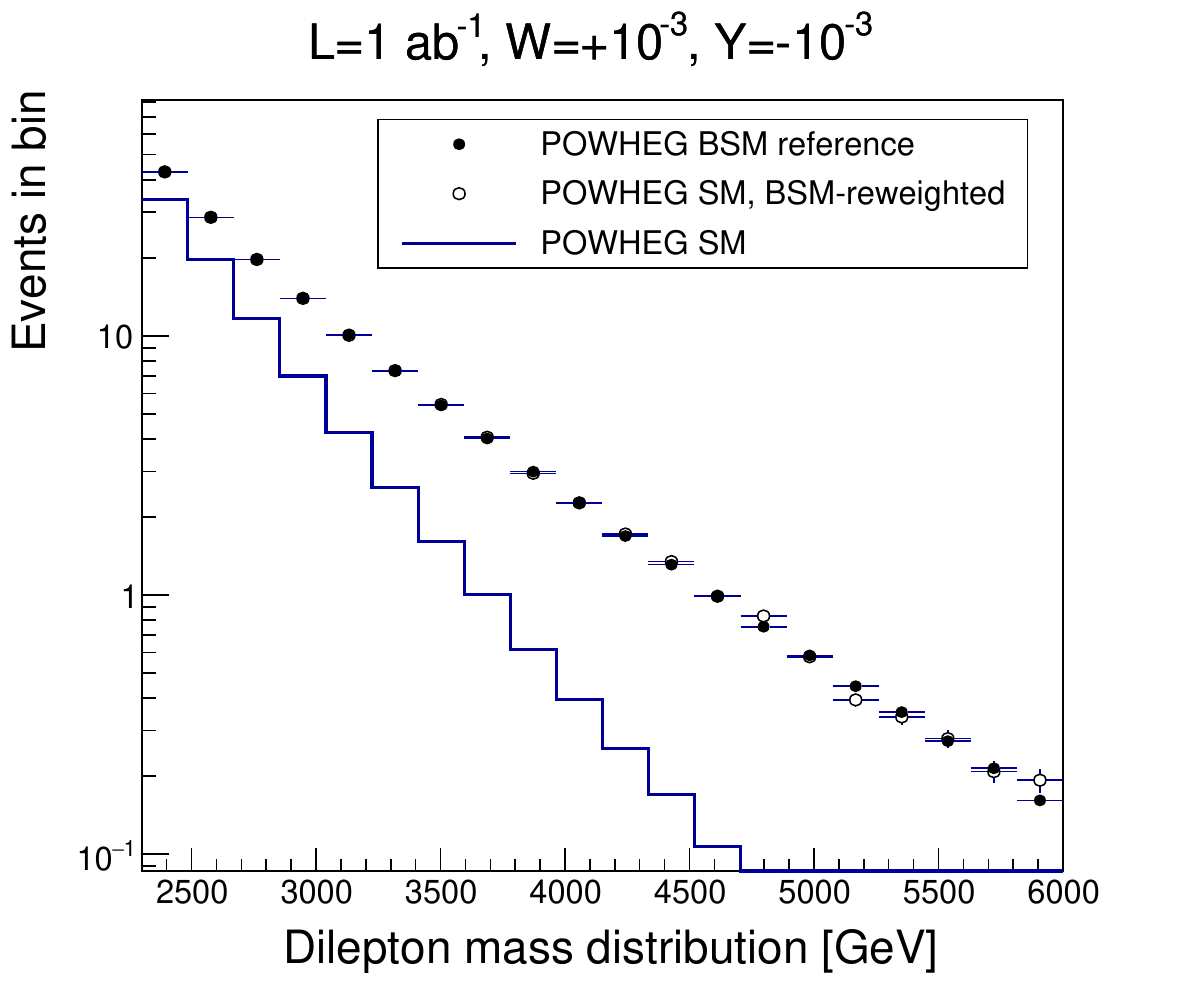}
\includegraphics[width=0.45\linewidth]{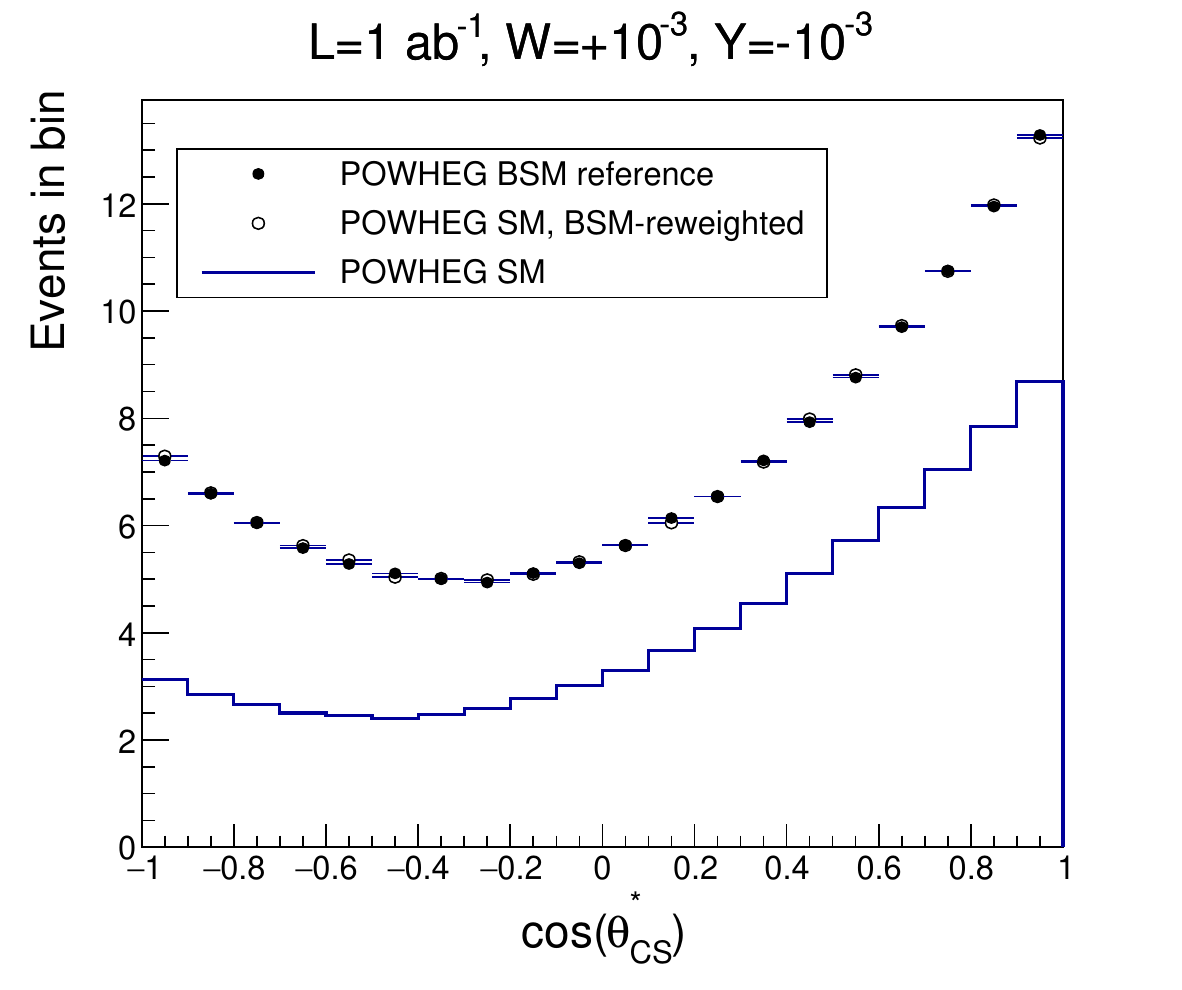}

\caption{Mass and angular comparisons between a BSM POWHEG sample generated with
$W=+10^{-3}, Y=-10^{-3}$ and the corresponding result from the new proposed reweighting on
a pure SM sample inclusive in helicities. The samples have been generated with invariant
masses of the dilepton system in the intervals $1.4 < M_{\llbar} < 2.3~\TeV$ and $2.3 <
M_{\llbar} < 6.0~\TeV$ at $\sqrt{s}=13~\TeV$, for an integrated luminosity of 
${\cal L}=1~\abinv$. The SM expectations are also shown to illustrate the sensitivity to 
BSM effects.}

\label{fig:BSM-pm}

\end{figure}

\begin{figure}[htbp]
\centering

\includegraphics[width=0.45\linewidth]{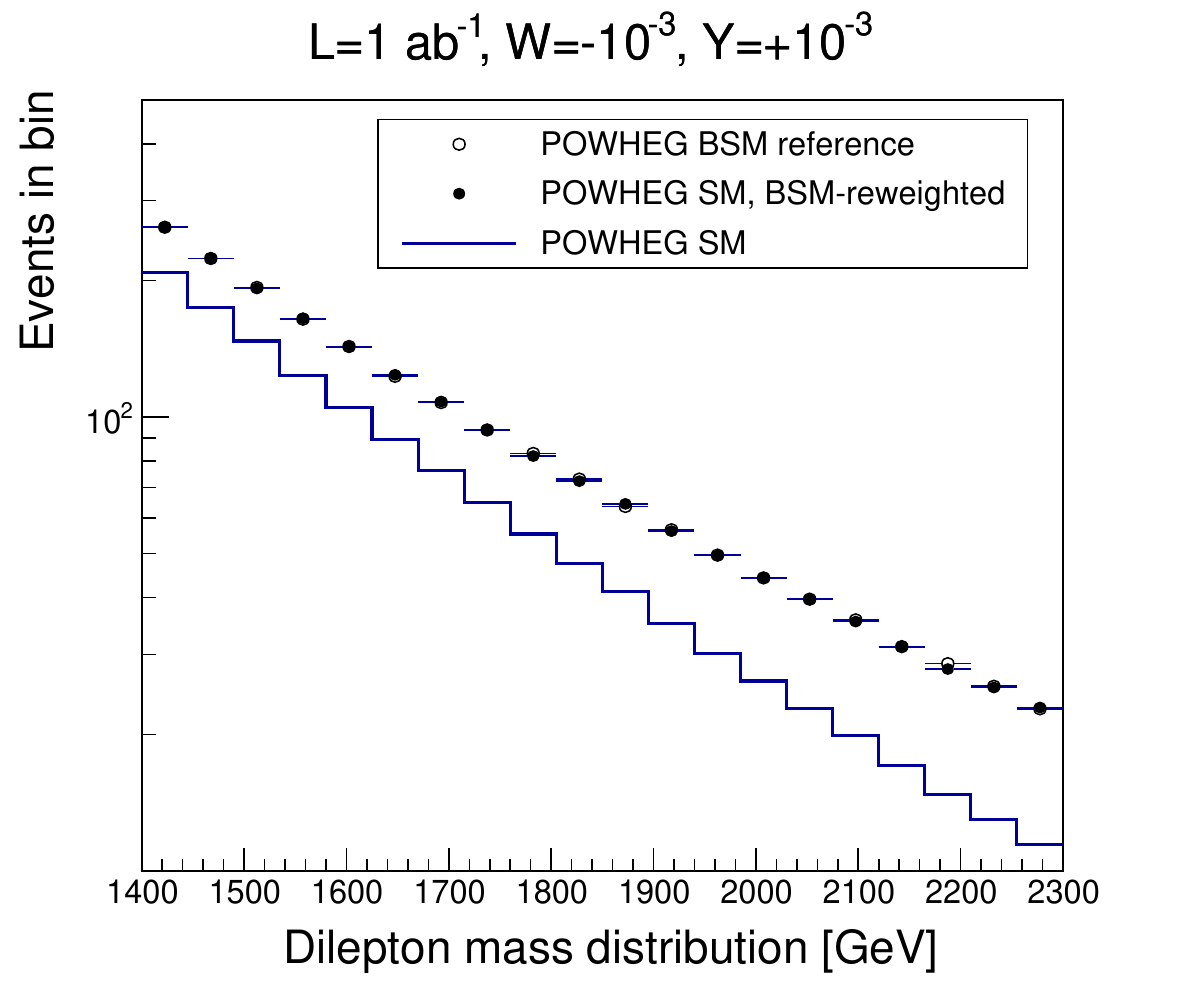}
\includegraphics[width=0.45\linewidth]{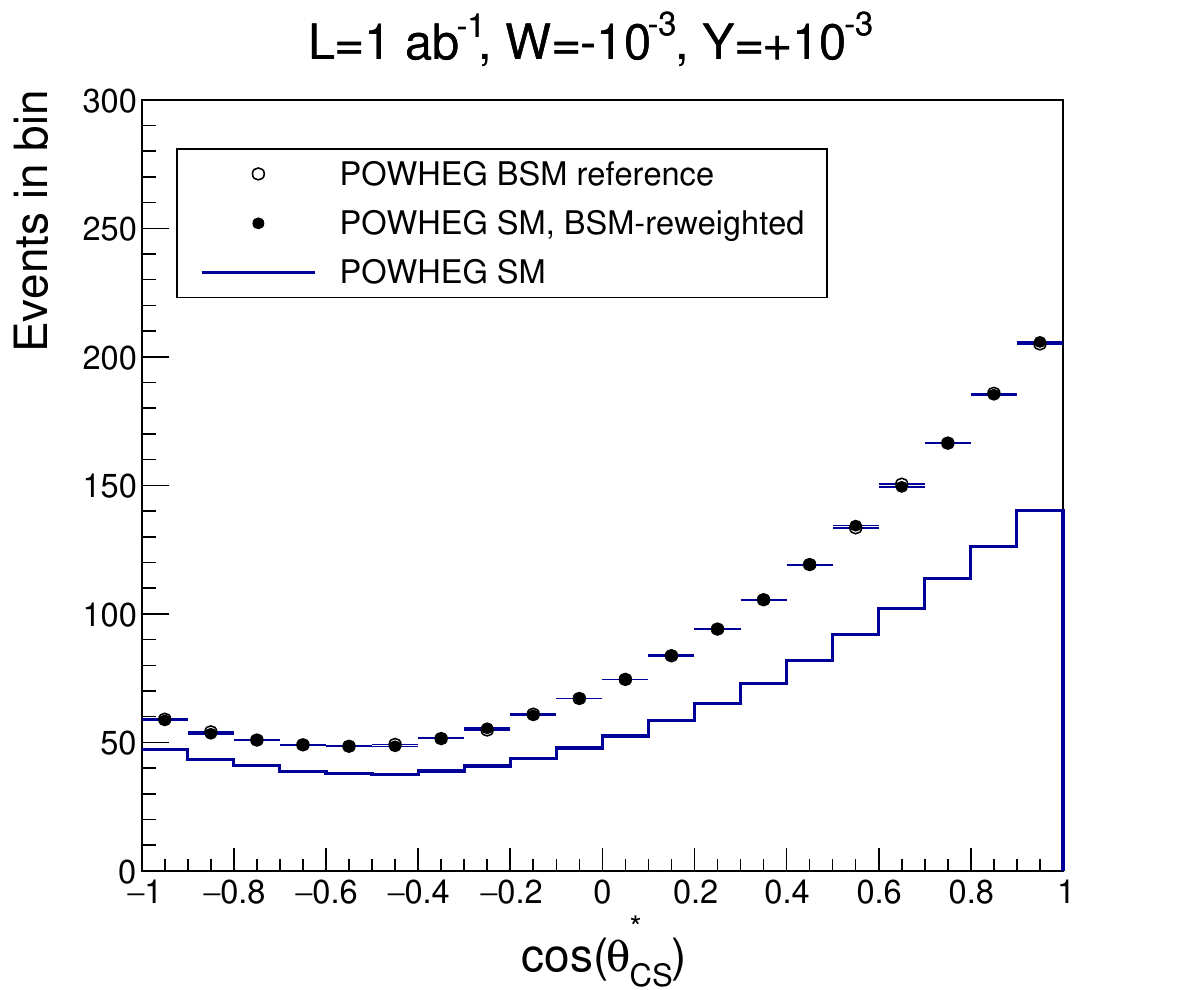}\\
\includegraphics[width=0.45\linewidth]{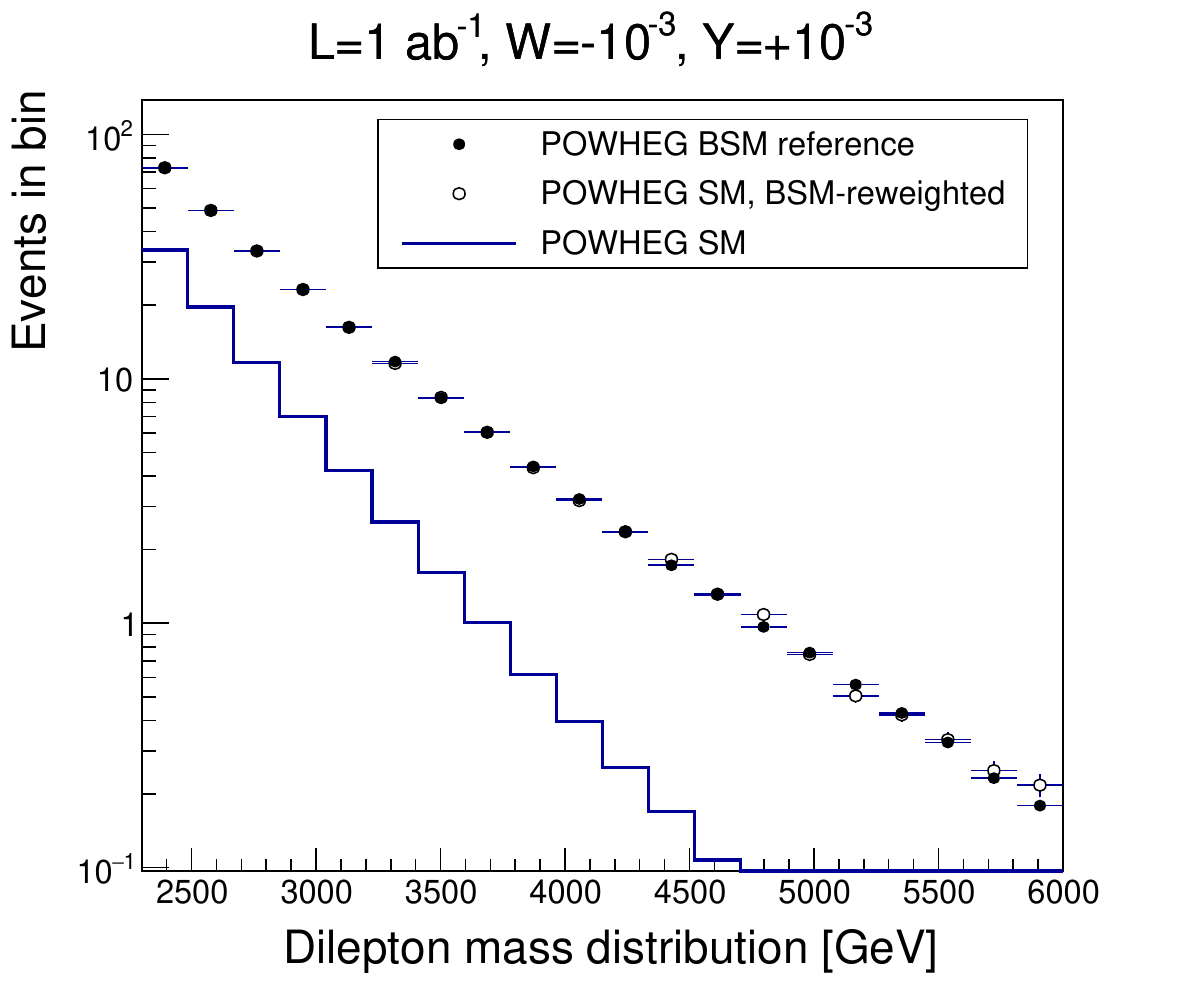}
\includegraphics[width=0.45\linewidth]{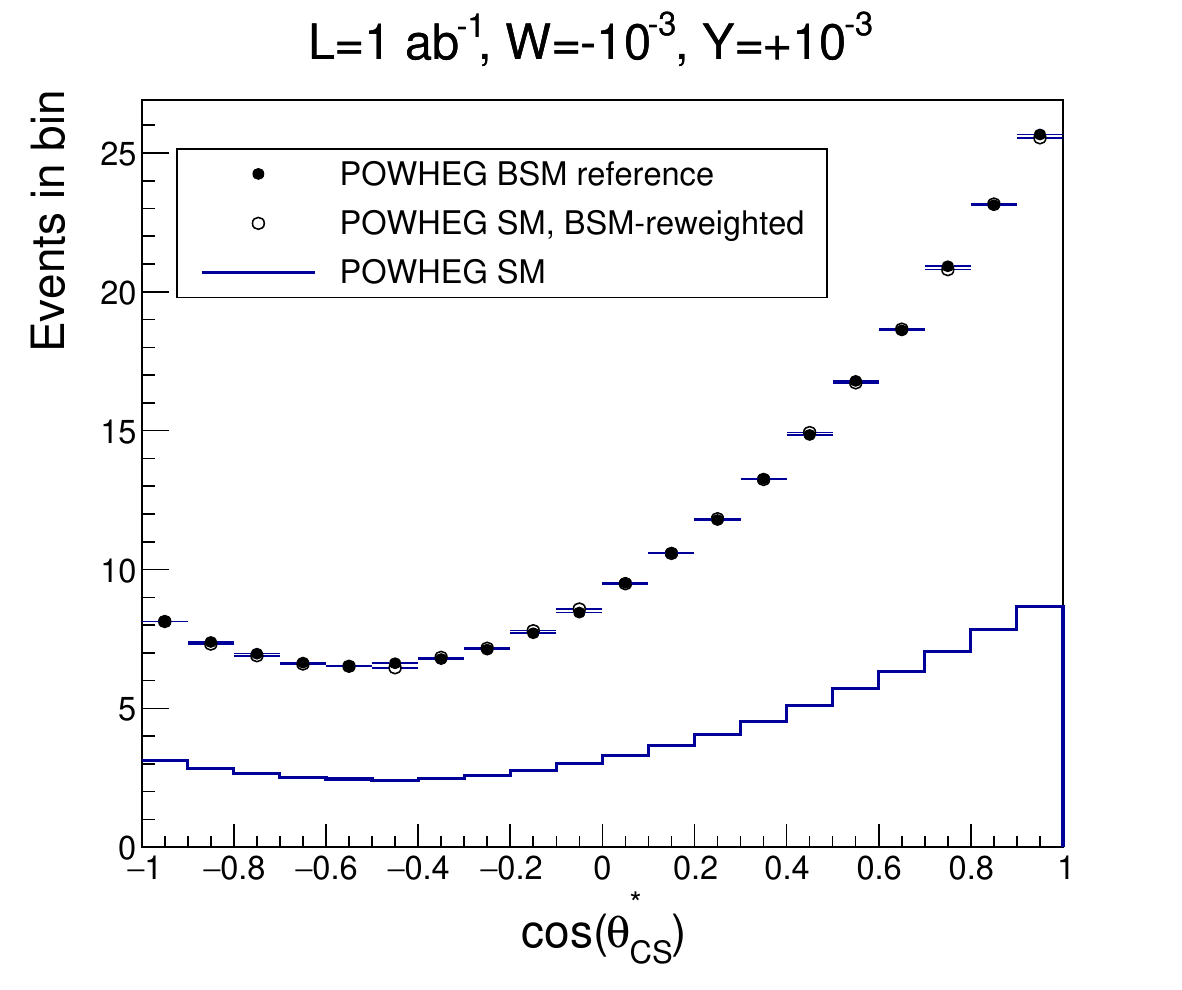}

\caption{Mass and angular comparisons between a BSM POWHEG sample generated with
$W=-10^{-3}, Y=+10^{-3}$ and the corresponding result from the new proposed reweighting on
a pure SM sample inclusive in helicities. The samples have been generated with invariant
masses of the dilepton system in the intervals $1.4 < M_{\llbar} < 2.3~\TeV$ and $2.3 <
M_{\llbar} < 6.0~\TeV$ at $\sqrt{s}=13~\TeV$, for an integrated luminosity of 
${\cal L}=1~\abinv$. The SM expectations are also shown to illustrate the sensitivity to 
BSM effects.}

\label{fig:BSM-mp}

\end{figure}

We also show in Figures~\ref{fig:BSM-pp}-\ref{fig:BSM-mp} the results for alternative BSM
scenarios. Reference and reweighted distributions are always found to be in
agreement. The scenario $\Wpar=\Ypar=+10^{-3}$, shown in Figure~\ref{fig:BSM-pp},
corresponds to a BSM cross section much smaller than the one in the SM. Note the
significant decrease in cross section around $M_{\llbar}\approx
\MW/\sqrt{\Wpar}\equiv\MW/\sqrt{\Ypar}$, which is expected due to the similarities between
BSM and SM structures at high mass. Figure~\ref{fig:BSM-pm} illustrates a scenario
($\Wpar=+10^{-3}, \Ypar=-10^{-3}$) where the number of events for the SM and BSM hypotheses
are more similar. There, mass and angular shapes may become critical to establish which 
model is preferred by the data. Finally, Figure~\ref{fig:BSM-mp} ($\Wpar=-10^{-3}, 
\Ypar=+10^{-3}$) shows an intermediate scenario, qualitatively similar to 
$\Wpar=\Ypar=-10^{-3}$, but where the BSM excess is less pronounced. In all cases, as 
expected, there is a significant BSM increase at the highest masses, proportional to the 
fourth power of the dilepton invariant mass.

\begin{figure}[htbp]
\centering
\includegraphics[width=0.45\linewidth]{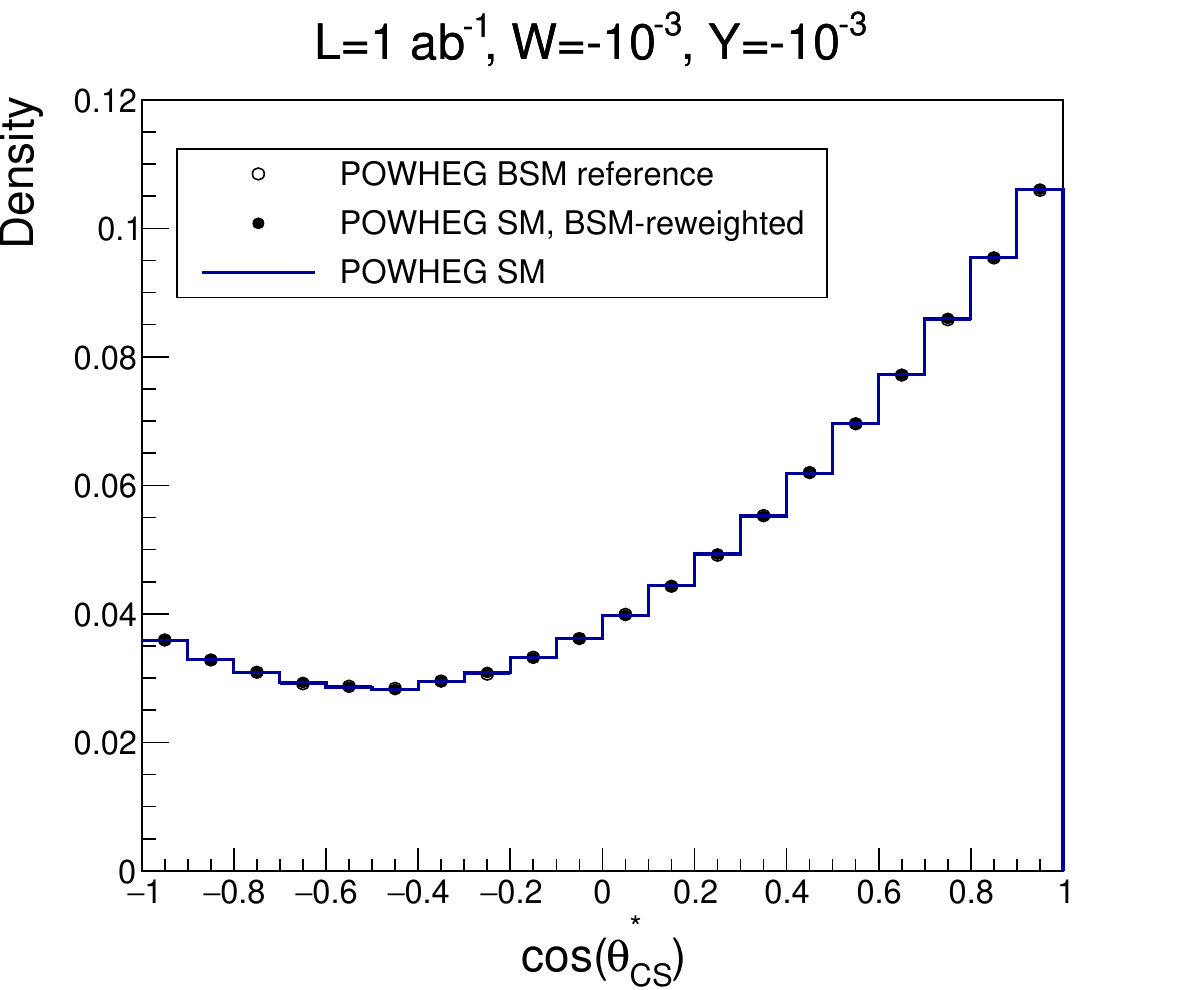}
\includegraphics[width=0.45\linewidth]{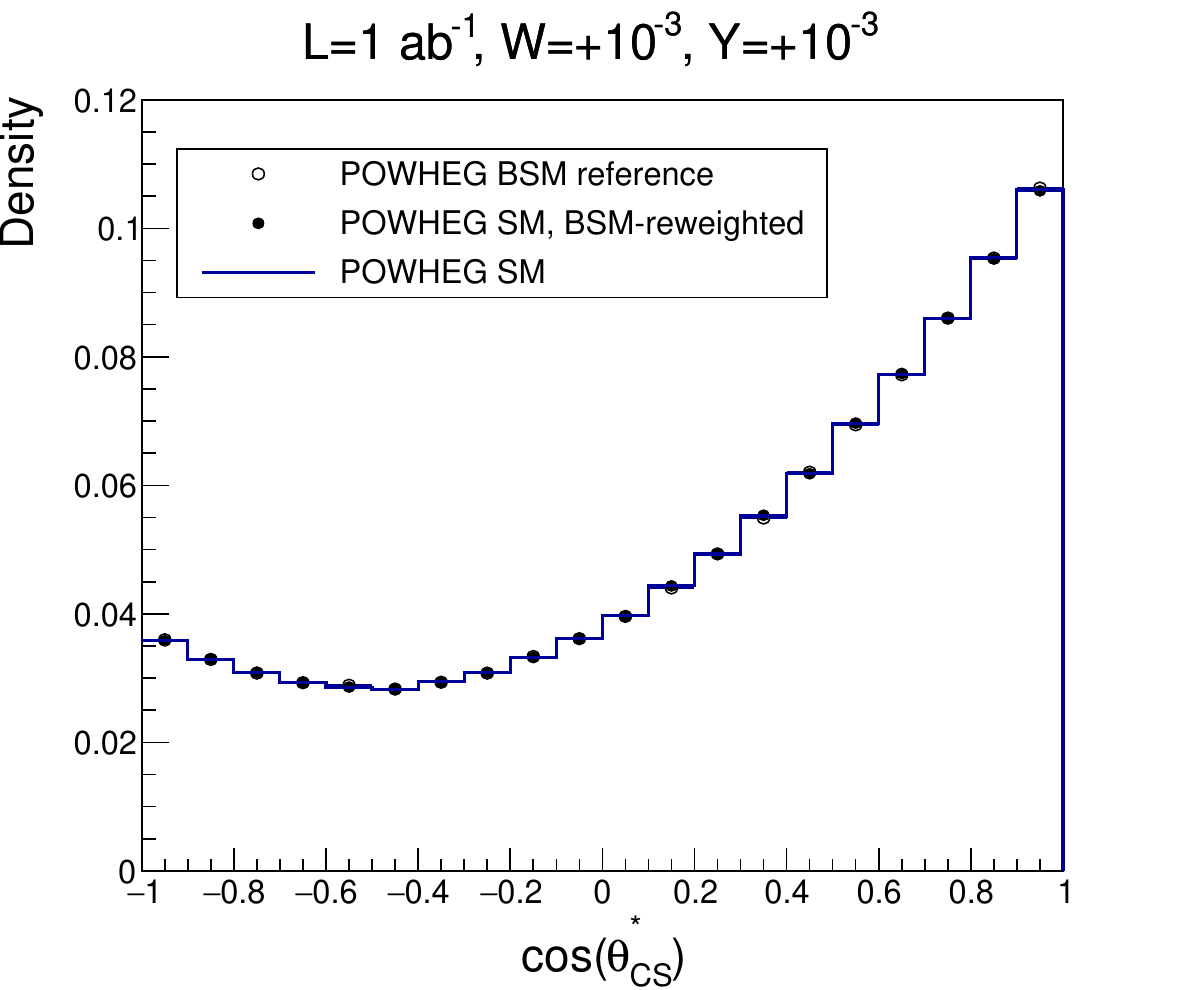}\\
\includegraphics[width=0.45\linewidth]{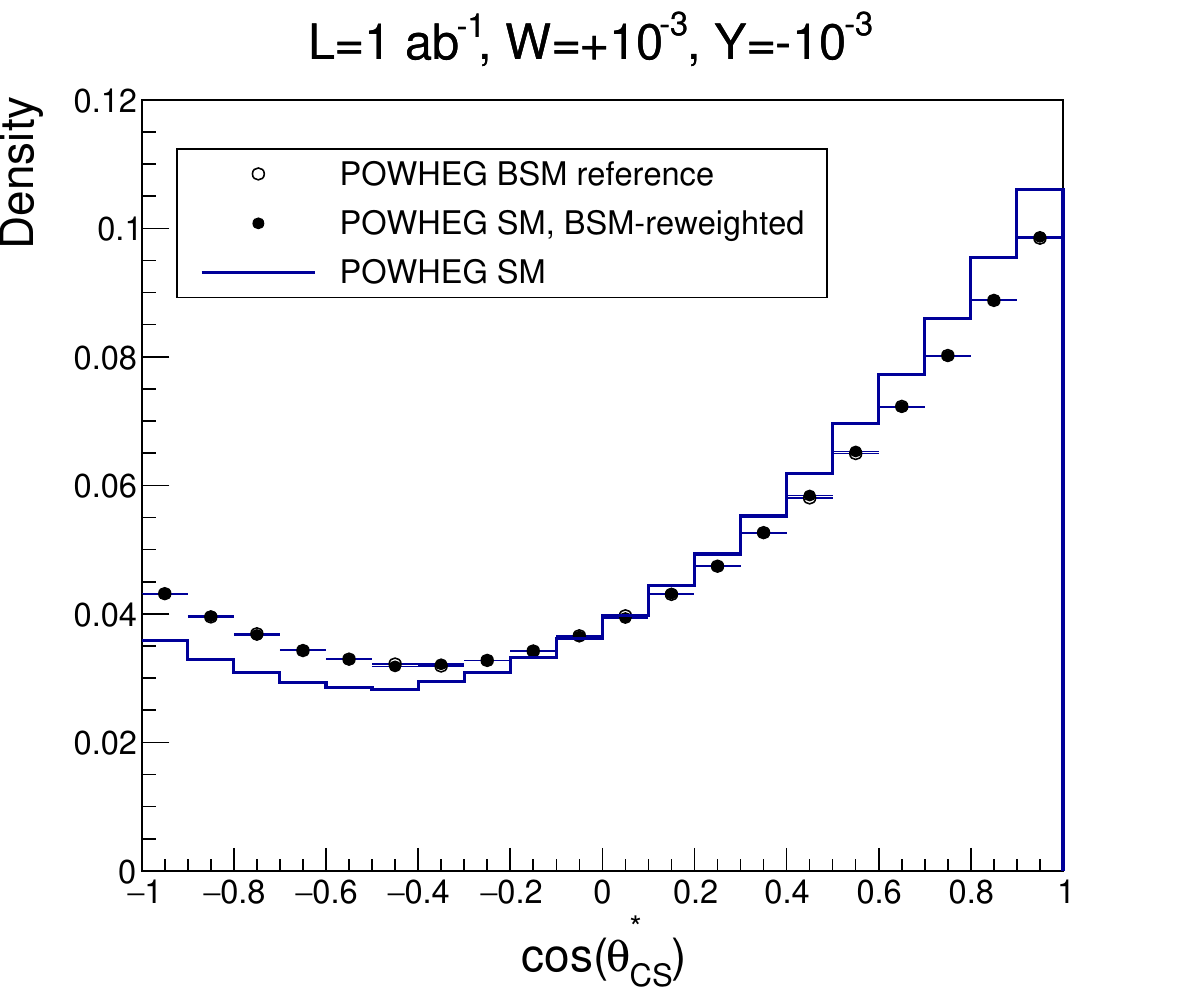}
\includegraphics[width=0.45\linewidth]{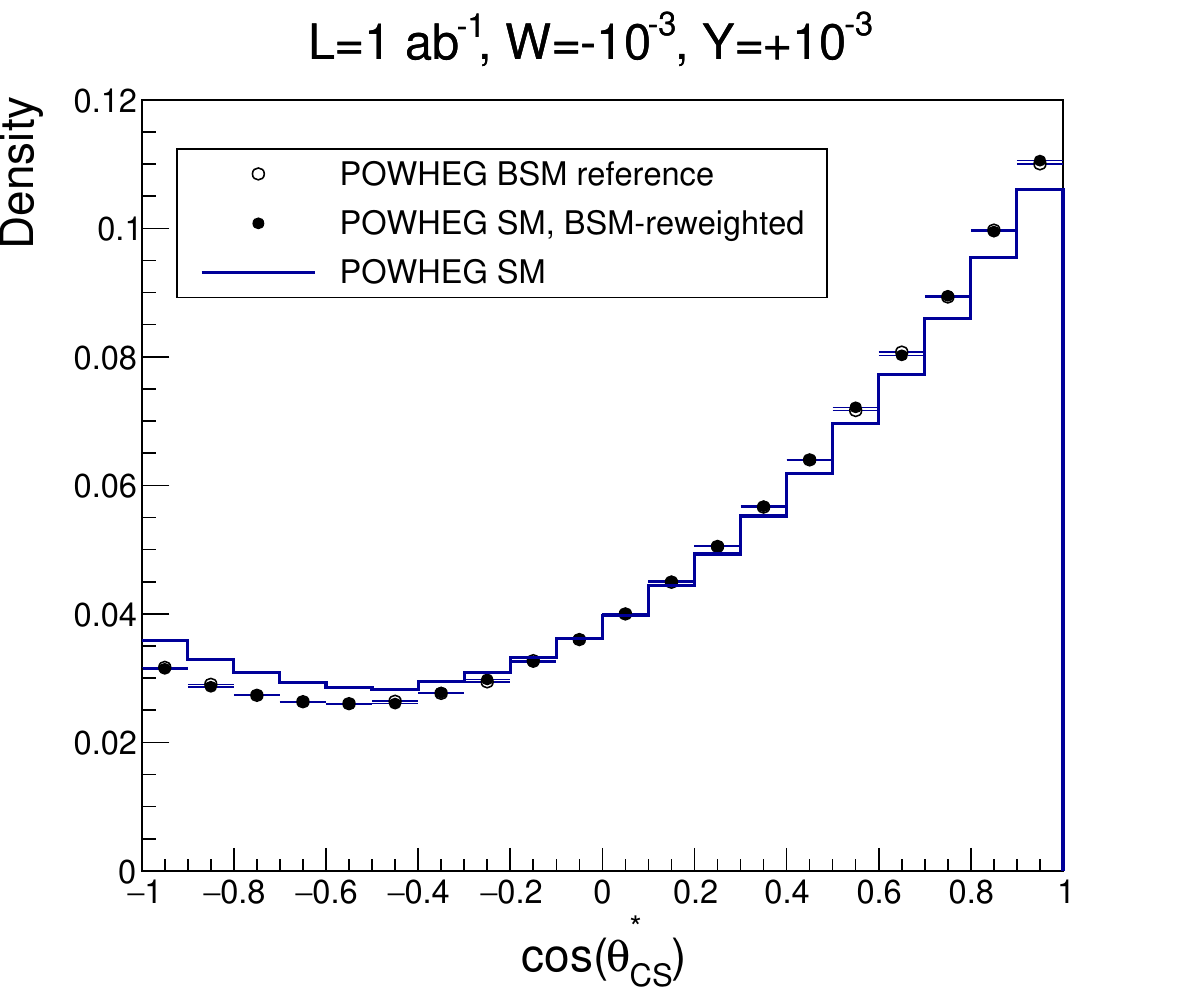}

\caption{Normalized angular distributions for different values of the $\Wpar,\Ypar$ oblique
parameters in the $1.4 < M_{\llbar} < 2.3~\TeV$ mass range. Only scenarios with different
values of \Wpar and \Ypar show visible differences with respect to the SM behaviour.
\label{fig:AFB}}

\end{figure}

\begin{figure}[htbp]
\centering
\includegraphics[width=0.45\linewidth]{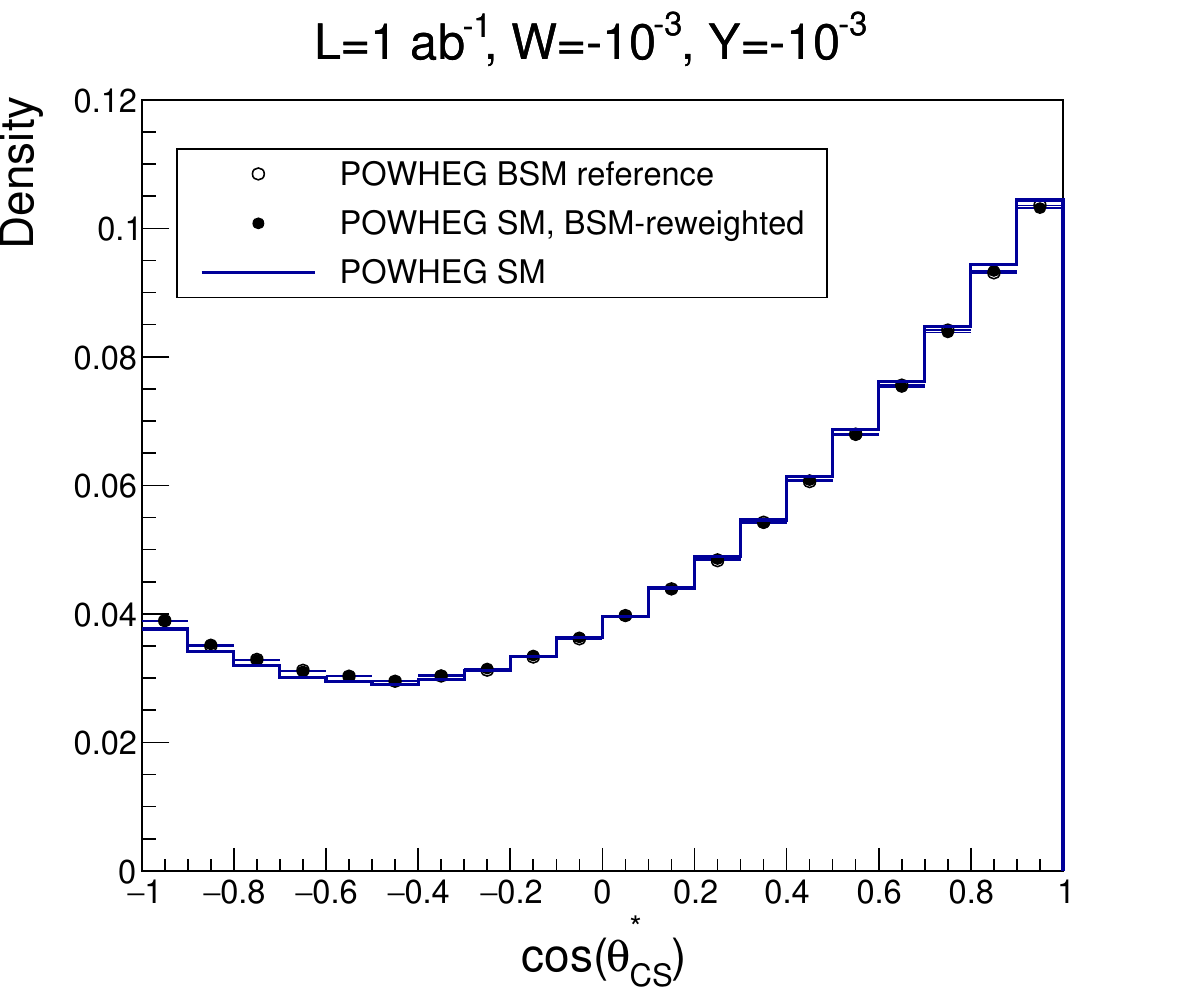}
\includegraphics[width=0.45\linewidth]{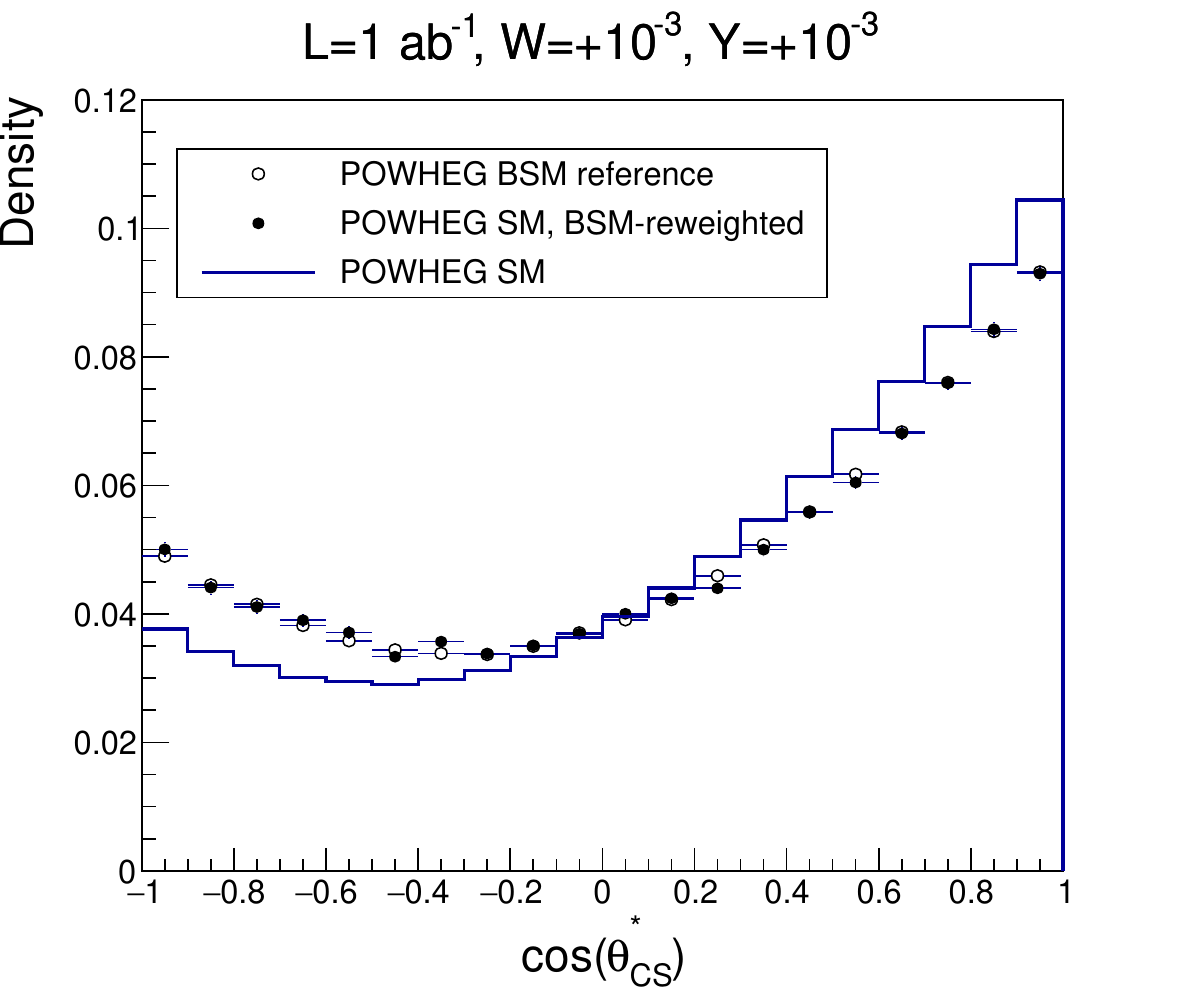}\\
\includegraphics[width=0.45\linewidth]{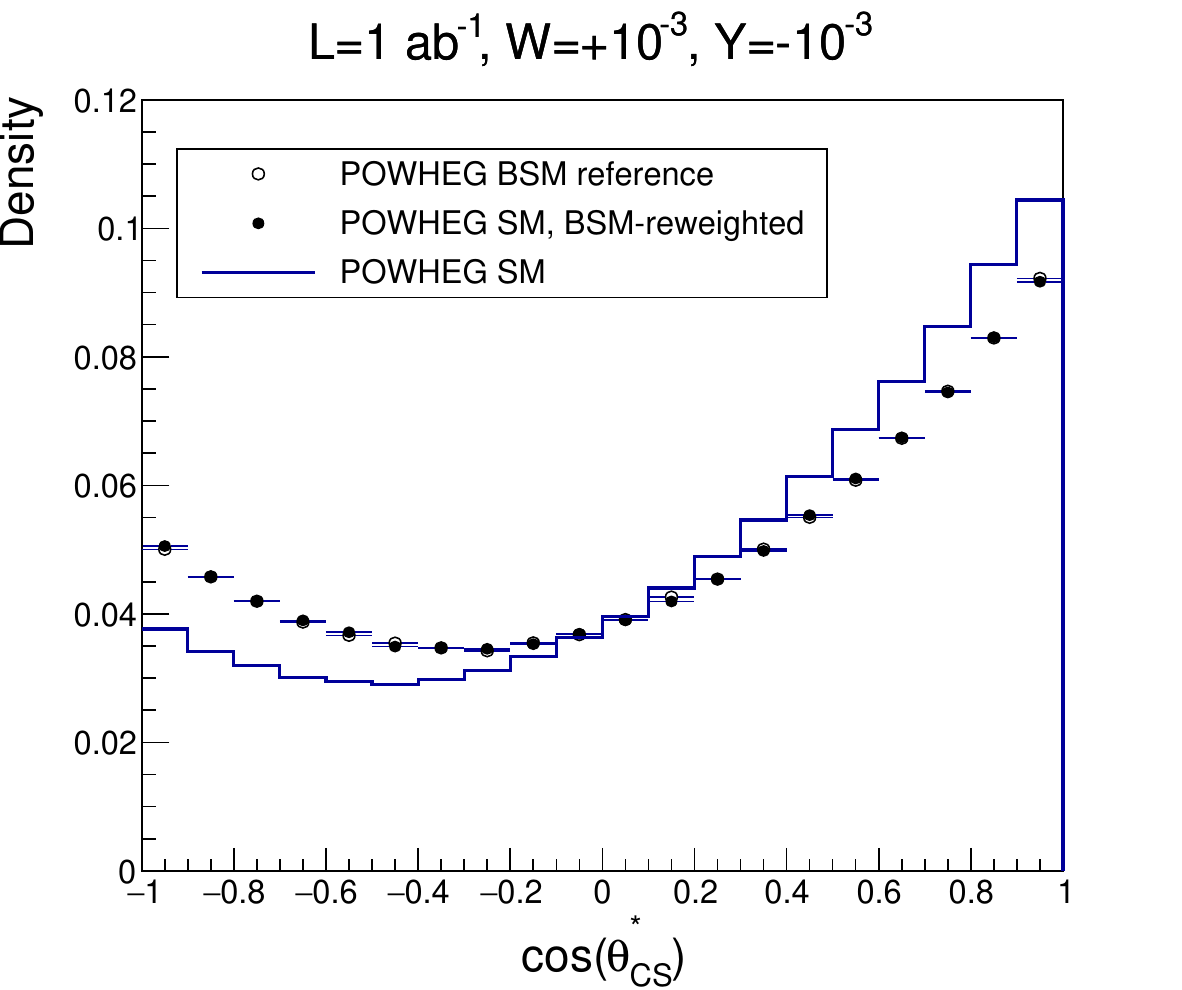}
\includegraphics[width=0.45\linewidth]{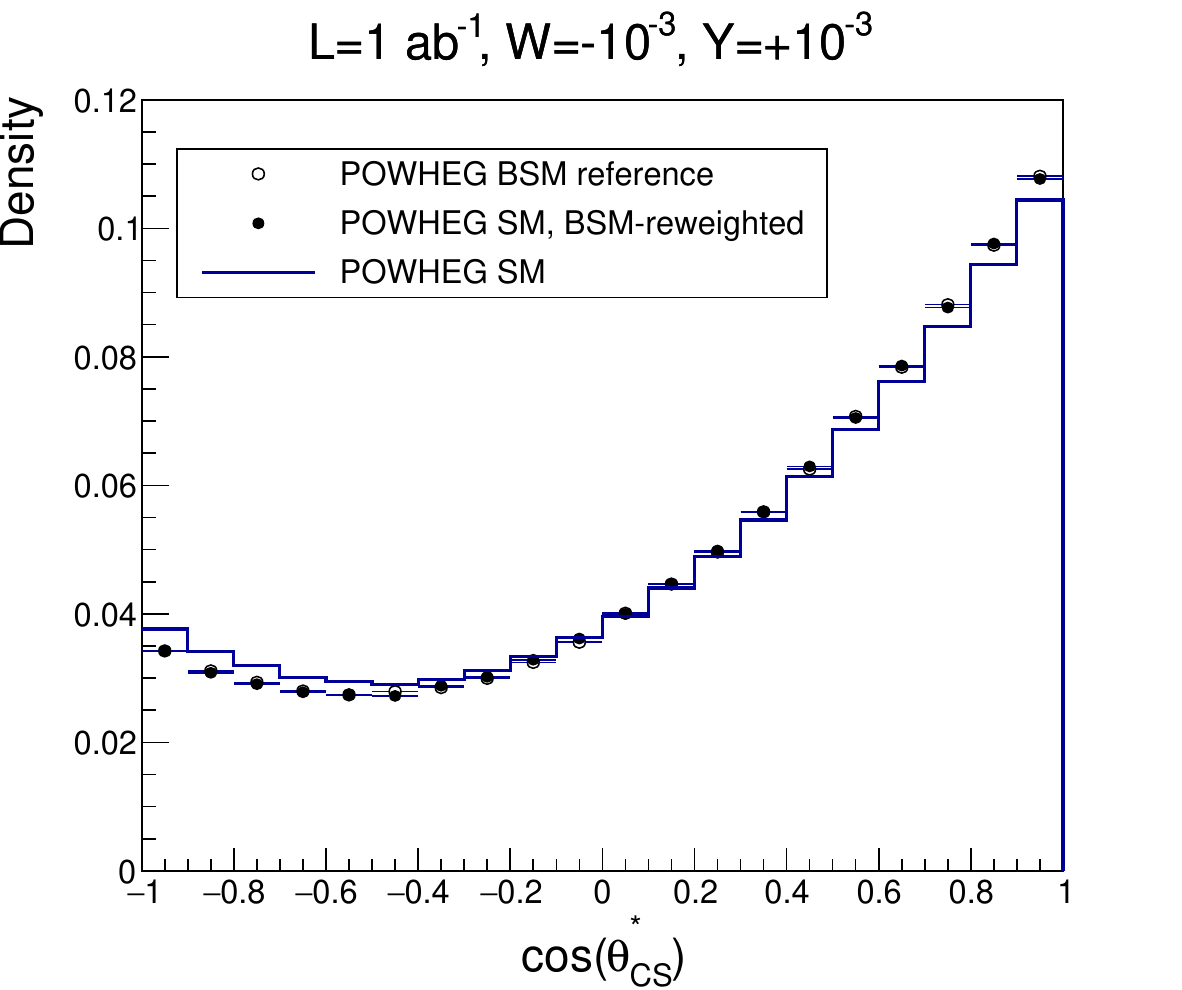}

\caption{Normalized angular distributions for different values of the $\Wpar,\Ypar$ oblique parameters  in the $2.3 < M_{\llbar} < 6.0~\TeV$ mass range. \label{fig:AFBhigh}}

\end{figure}

As a byproduct we have also determined the observable asymmetries in those BSM scenarios,
$\AFB^* \equiv \frac{N(\cos\theta^*_{CS}>0)-N(\cos\theta^*_{CS}<0)}
{N(\cos\theta^*_{CS}>0)+N(\cos\theta^*_{CS}<0)}$, for the two mass ranges under study.
These asymmetries are shown in Table~\ref{tab:angular-BSM}. The reference and reweighted
calculations of $\AFB^*$ agree within the estimated statistical uncertainties. In the
absence of BSM effects, the estimated asymmetries in the SM for the $1.4 < M_{\llbar} <
2.3~\TeV$ and $2.3 < M_{\llbar} < 6.0~\TeV$ mass ranges are $\AFB^*(SM) = (36.93 \pm
0.07)\%$ and $\AFB^*(SM) = (36.93 \pm 0.07)\%$, respectively. Note that, in the first mass
range, the SM asymmetry is nearly equal to the BSM asymmetry for models with $\Wpar=\Ypar$.
This is expected because the differential cross section for a BSM model with $\Wpar=\Ypar$
is almost exactly proportional to the SM one in this range. This is more clearly seen in
the normalized angular distributions of Figure~\ref{fig:AFB}. The situation is rather
different for masses above $2.3~\TeV$, as shown in Figure~\ref{fig:AFBhigh}. This is due to
the fact that the chosen values of $\Wpar$ and $\Ypar$ imply new physics at a scale that is
closer to the explored range, where the differential cross sections start to differ
significantly.

\begin{table}[htb]
\begin{center}

\begin{tabular}{|c|c|c|} \hline
BSM scenario, $1.4 < M_{\llbar} < 2.3~\TeV$ & $\AFB^*$ reference & $\AFB^*$ from reweighting procedure \\
\hline
$\Wpar=-10^{-3},\Ypar=-10^{-3}$ & $(36.89 \pm 0.07)\%$ & $(36.94 \pm 0.07)\%$ \\
$\Wpar=+10^{-3},\Ypar=+10^{-3}$ & $(36.84 \pm 0.08)\%$ & $(36.91 \pm 0.08)\%$ \\
$\Wpar=+10^{-3},\Ypar=-10^{-3}$ & $(29.10 \pm 0.07)\%$ & $(29.18 \pm 0.08)\%$ \\
$\Wpar=-10^{-3},\Ypar=+10^{-3}$ & $(41.48 \pm 0.07)\%$ & $(41.63 \pm 0.07)\%$ \\
\hline
\end{tabular}

\vspace{0.5cm}

\begin{tabular}{|c|c|c|} \hline
BSM scenario, $2.3 < M_{\llbar} < 6.0~\TeV$ & $\AFB^*$ reference & $\AFB^*$ from reweighting procedure \\
\hline
$\Wpar=-10^{-3},\Ypar=-10^{-3}$ & $(34.22 \pm 0.07)\%$ & $(34.07 \pm 0.08)\%$ \\
$\Wpar=+10^{-3},\Ypar=+10^{-3}$ & $(23.41 \pm 0.07)\%$ & $(22.77 \pm 0.44)\%$ \\
$\Wpar=+10^{-3},\Ypar=-10^{-3}$ & $(22.03 \pm 0.07)\%$ & $(21.82 \pm 0.11)\%$ \\
$\Wpar=-10^{-3},\Ypar=+10^{-3}$ & $(39.09 \pm 0.07)\%$ & $(39.06 \pm 0.09)\%$ \\
\hline
\end{tabular}

\caption{Comparison of the forward-backward asymmetries $\AFB^*$ as a function of the
observed Collins-Soper polar angle $\cos\theta^*_{CS}$ between SM POWHEG samples generated
in BSM scenarios with different values of the $\Wpar,\Ypar$ oblique parameters and the
results of the new proposed reweighting on a pure SM sample inclusive in helicities.
The samples have been generated at $\sqrt{s}=13~\TeV$ in the dilepton invariant mass ranges
$1.4 < M_{\llbar} < 2.3~\TeV$ and $2.3 < M_{\llbar} < 6.0~\TeV$. In the absence of BSM 
effects, the expected asymmetries (SM) are $\AFB^* = (36.93 \pm 0.07)\%$ and 
$\AFB^* = (35.39 \pm 0.07)\%$, respectively.}

\label{tab:angular-BSM}

\end{center}
\end{table}

\section{Other physics scenarios that can be targeted by the procedure}

As commented before, a scenario identified by a set of $\Wpar,\Ypar$ oblique
parameters can be interpreted in terms of new contact interactions at a scale of new physics
$\Lambda$ ($\Lambda\gg \sqrt{s}$)~\cite{Andrea-main,CMS-Run2-ll}:

\begin{eqnarray}
	-\frac{g^2~\Wpar~\tq~\tl}{\MW^2}, -\frac{g^{\prime 2}~\Ypar~\yq~\yl}{\MW^2} & ~~\longrightarrow~~ & \frac{c_{\chiq\chil}}{\Lambda^2}\;.
	\label{eq:CI-again}
\end{eqnarray}

Let us nevertheless point out some differences between contact interactions in the context
of Higgs compositeness and usual searches for $\qqbar\llbar$ contact interactions at the
LHC. First, the size of the coupling in usual LHC searches is chosen to be as large as
possible, at the unitarity boundary: $c_{\chiq\chil}(LHC)\equiv \sqrt{4\pi}$. This ensures
that $\Lambda \gg \sqrt{s}$, and therefore stringent lower limits on $\Lambda$, although it
has little physical motivation. Another subtle point is that, for the $LL$ case, usual
searches target the $Q^{(1)}_{\ell\Pq}$ operator of the Warsaw basis~\cite{Warsaw}, while
in Higgs compositeness the ${\cal O}'_{2W}$ operator is of the $Q^{(3)}_{\ell\Pq}$ type.
From a practical point of view, this means that in Higgs compositeness studies there are
partially canceling BSM effects between up and down quarks, due to the presence of 
$\tq$ factors in that operator. 

Apparent evidence of contact interactions may, in reality, be a misinterpretation of the
presence of a true $\Zprime$ resonance with a relatively large width at the visible tail of
the mass spectrum. Ultimately, and given the high-tail statistics that will be available at
the HL-LHC, all scenarios must be considered and properly simulated. For this purpose, the
previous expressions can easily be adapted to the resonant case. Assuming the exchange of a
new $\Zprime$ of mass $M_{\Zprime}$ and chiral couplings $c_{\chiq}$ and $c_{\chil}$ to
quarks and leptons:: 

\begin{eqnarray}
	-\frac{g^2~\Wpar~\tq~\tl}{\MW^2}, -\frac{g^{\prime 2}~\Ypar~\yq\yl}{\MW^2} & ~~\longrightarrow~~ & \frac{c_{\chiq}~c_{\chil}}{s-M^2_{\Zprime}+i\Gamma_{\Zprime} M_{\Zprime}}\;,
\end{eqnarray}

\noindent where $\Gamma_{\Zprime}$ is the width of the resonance. Note that in the $\Lambda
\equiv M_{\Zprime} \gg \sqrt{s}$ limit one recovers the contact interaction case presented in
Equation~\ref{eq:CI-again}, with $c_{\chiq\chil}\equiv -c_{\chiq}~c_{\chil}$. 

Finally, let us also note the proposed strategy can be extended without any concern to
non-universal BSM scenarios affecting only specific types of quarks and leptons.

\section{$\Zprime$ resonances in HVT and HVS models}

Let us first assume the Heavy Vector Triplet (HVT) model proposed in
Ref.~\cite{HVT}, with $c_f=1$, associated with a compositeness parameter $g^*\equiv g_V$ 
and a compositeness scale $m^*\equiv M_{\Zprime}$:

\begin{eqnarray}
	-\frac{g^2~\Wpar~\tq~\tl}{\MW^2} & ~~\longrightarrow~~ & \left(\frac{g^2}{g^{*}}\right)^2~\left(\frac{\tq~\tl}{s-M^2_{\Zprime}+i\Gamma_{\Zprime} M_{\Zprime}}\right)\;,
\end{eqnarray}

\noindent where:

\begin{eqnarray}
	\frac{\Gamma_{\Zprime}}{M} & = & \frac{g^{*2}}{96\pi} + \sum_{fermions}\left(\frac{g^2}{g^{*}}\right)^2~\frac{N_c(f)~[\tf]^2}{24\pi} = \frac{g^{*2}}{96\pi} + \frac{g^4}{4\pi~g^{* 2}} \;,
\end{eqnarray}

\noindent being $N_c(f)$ is the number of colors for the fermion $f$. Note that the total
width $\Gamma_{\Zprime}$ includes a partial width fraction into WW and ZH, which is
proportional to $g^{*2}$. 

\begin{figure}[htbp]
\centering

\includegraphics[width=0.45\linewidth]{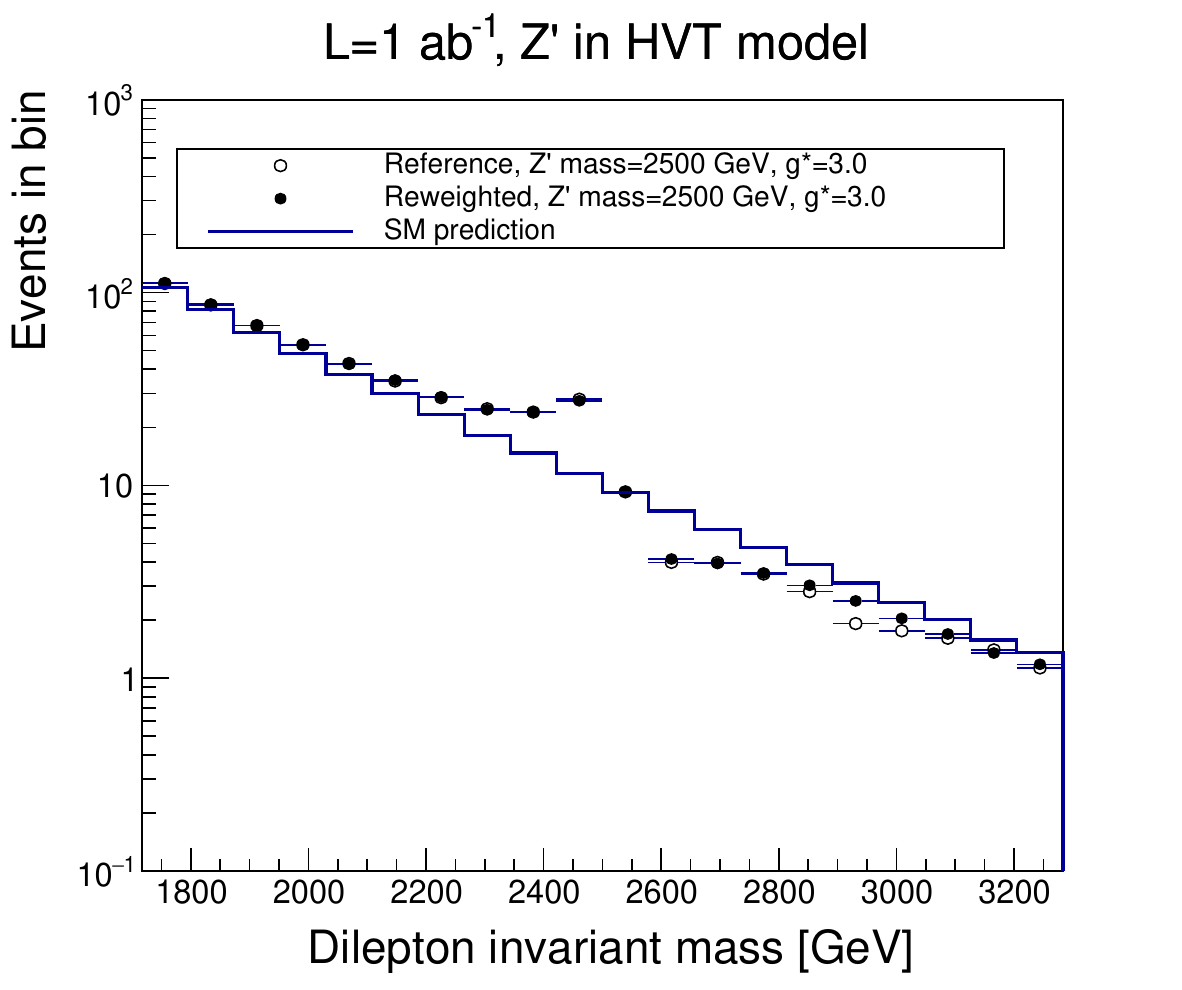}
\includegraphics[width=0.45\linewidth]{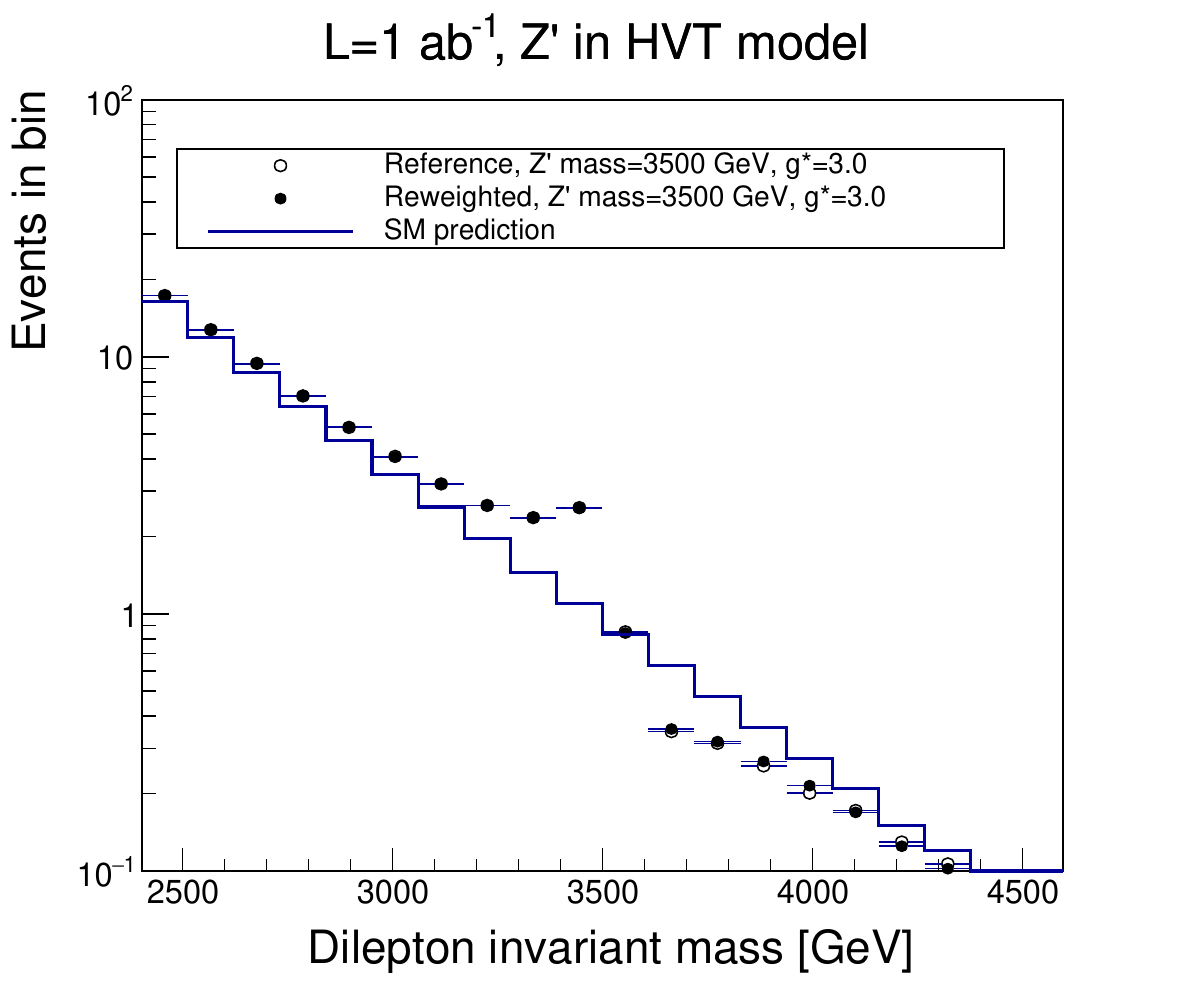}

\caption{Mass comparisons between BSM POWHEG HVT samples generated with
$m_{\Zprime}=2.5~\TeV, g^*=3$ and $m_{\Zprime}=3.5~\TeV, g^*=3$, and the corresponding
result from the new proposed reweighting on a pure SM sample inclusive in helicities. The
samples have been generated at $\sqrt{s}=13~\TeV$, for an integrated luminosity of ${\cal
L}=1~\abinv$. The SM expectations are also shown to illustrate the sensitivity to BSM
effects.} 

\label{fig:HVT}

\end{figure}

Figure~\ref{fig:HVT} shows two examples of comparisons between the POWHEG-generated HVT
$\Zprime$ samples and reweighted ones. The distributions agree within the expected
uncertainties, despite the large weights that are present at some invariant masses. Note
that the signal shapes can be rather complicated due to the small size of the signal, the
interference with the SM, and PDF effects.

Similarly, we can consider a Y-sequential case, of the Heavy Vector Singlet (HVS)~\cite{HVS} type, associated with a  compositeness parameter $g^*$ and a compositeness scale 
$m^*\equiv M_{\Zprime}$:

\begin{eqnarray}
	-\frac{g^{\prime 2}~\Ypar~\yq~\yl}{\MW^2} & ~~\longrightarrow~~ & \left(\frac{g^{\prime 2}}{g^{*}}\right)^2~\left(\frac{\yq~\yl}{s-M^2_{\Zprime}+i\Gamma_{\Zprime} M_{\Zprime}}\right)\;,
\end{eqnarray}

\noindent where now:

\begin{eqnarray}
	\frac{\Gamma_{\Zprime}}{M} & = & \frac{g^{*2}}{96\pi} + \sum_{fermions}\left(\frac{g^{\prime 2}}{g^{*}}\right)^2~\frac{N_c(f)~[\yf]^2}{24\pi} = \frac{g^{*2}}{96\pi} + \frac{5~g^{\prime 4}}{12\pi~g^{* 2}} \;.
\end{eqnarray}

\begin{figure}[htbp]
\centering

\includegraphics[width=0.45\linewidth]{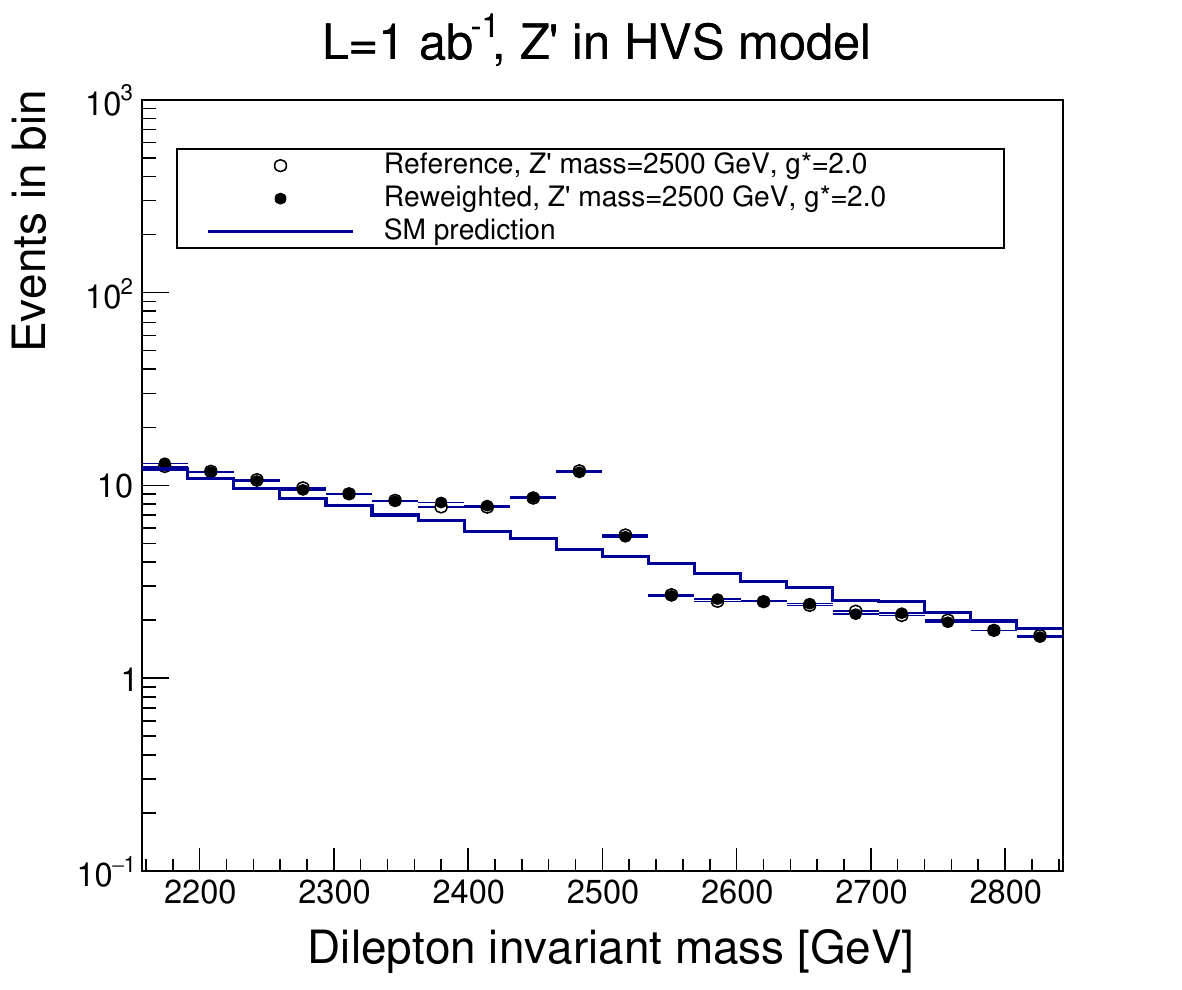}
\includegraphics[width=0.45\linewidth]{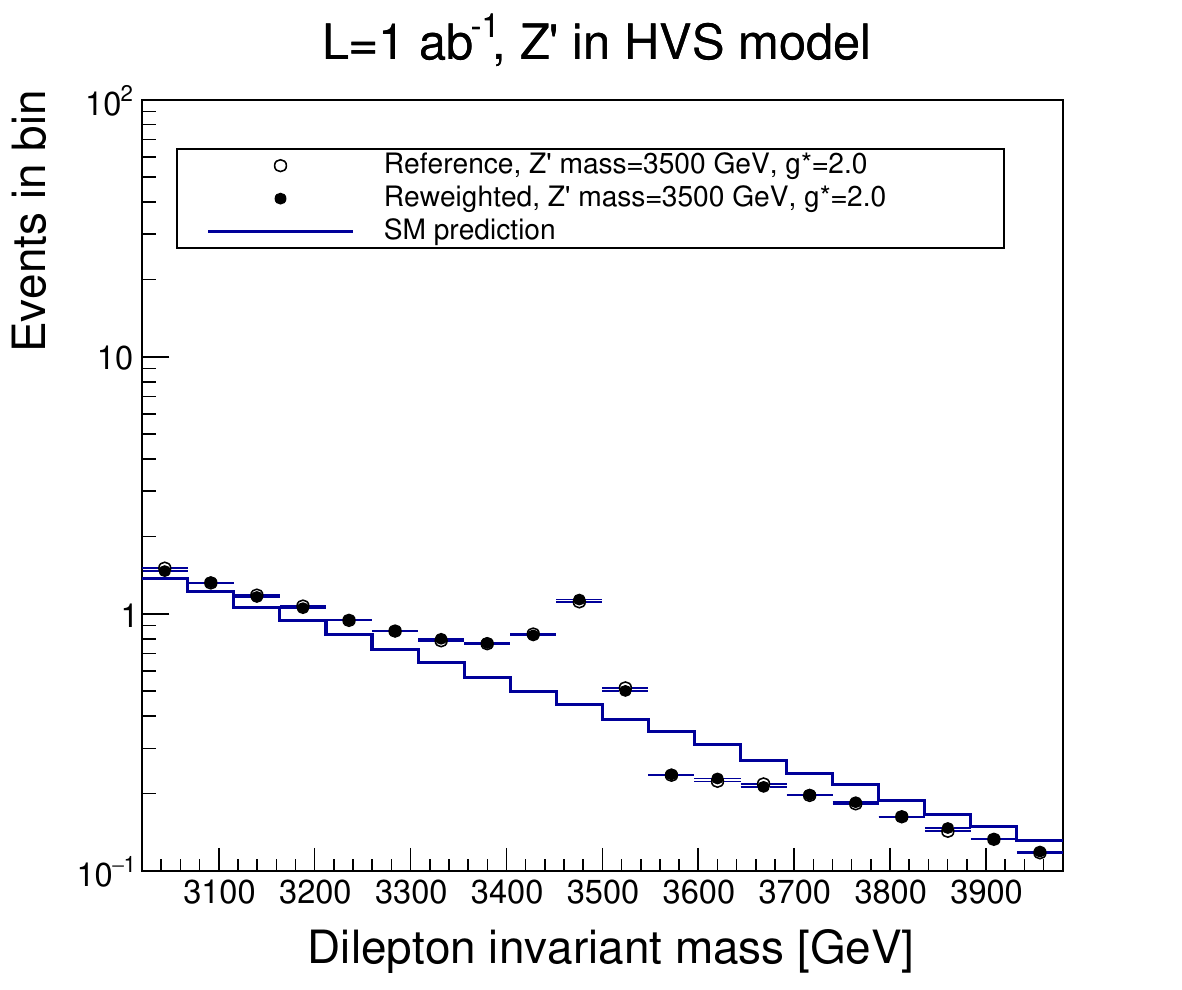}

\caption{Mass comparisons between BSM POWHEG HVS samples generated with
$m_{\Zprime}=2.5~\TeV, g^*=2$ and $m_{\Zprime}=3.5~\TeV, g^*=2$, and the corresponding
result from the new proposed reweighting on a pure SM sample inclusive in helicities. The
samples have been generated at $\sqrt{s}=13~\TeV$, for an integrated luminosity of ${\cal
L}=1~\abinv$. The SM expectations are also shown to illustrate the sensitivity to BSM
effects.} 

\label{fig:HVS}

\end{figure}

Figure~\ref{fig:HVS} shows two more examples of comparisons between the POWHEG-generated
HVS $\Zprime$ samples and the reweighted ones. Like the HVT case, the distributions
are consistent within the expected uncertainties, with similar qualitative features.

\section{Summary}

We presented an optimal reweighting procedure for studying new physics effects in high-mass
dilepton production at the LHC with next-to-leading-order precision in quantum
chromodynamics. The method allows experimental analyses to explore arbitrary $\qqbar\llbar$
contact-interaction scenarios and medium-to-large width $\Zprime$ resonances, without
requiring computationally expensive simulations to cover all beyond-the-standard-model
parameter hypotheses. The approach preserves both normalization and shape accuracy in the
full phase space, making it well suited for future high-luminosity LHC searches.

\bibliographystyle{myutphys}
\bibliography{ZPRIME_note}{}

\end{document}